\documentclass[
  a4paper,
 superscriptaddress,
 amsmath,amssymb,
 reprint,
 aps,
 longbibliography,
 floatfix,
 notitlepage,
 showpacs,
]{revtex4-2}

\usepackage{soul,xcolor}
\usepackage{physics}
\usepackage{wrapfig, blindtext}
\usepackage{graphicx}
\usepackage{dcolumn}
\usepackage{bm}
\usepackage{verbatim}
\usepackage{cancel}
\usepackage{float}

\begin{document}

\title{Hybrid optomechanical superconducting qubit system}

\author{Juuso Manninen} \affiliation{Department of Science and
  Industry Systems, University of South-Eastern Norway, PO Box 235,
  Kongsberg, Norway}

\author{Robert H. Blick}
\affiliation{
 Center for Hybrid Nanostructures (CHyN), Universit\"at Hamburg,
Luruper Chaussee 149, Hamburg 22761 Germany}

\affiliation{
 Materials Science and Engineering, University of Wisconsin-Madison,
 1509 University Ave. WI 53706 U.S.A.}

\author{Francesco Massel} \email{francesco.massel@usn.no}
\affiliation{Department of Science and Industry Systems, University of
  South-Eastern Norway, PO Box 235, Kongsberg, Norway}

\date{\today}

\begin{abstract}
  We propose an integrated nonlinear superconducting device based 
  on a nanoelectromechanical shuttle. The system can be described
  as a qubit coupled to a bosonic mode. The topology of the circuit
  gives rise to an adjustable qubit/mechanical coupling, allowing the
  experimenter to tune between linear and quadratic coupling in the
  mechanical degrees of freedom. Owing to its flexibility and
  potential scalability, the proposed setup represents an
  important step towards the implementation of bosonic error
  correction with mechanical elements in large-scale superconducting
  circuits. We give preliminary evidence of this possibility by
  discussing a simple state-swapping protocol that uses this
  device as a quantum memory element.
\end{abstract}

\maketitle

%xxxxxxxxxxxxxxxxxxxxxxxxxxxxxxxx
% INTRODUCTION
%XXXXXXXXXXXXXXXXXXXXXXXXXXXXXXXX
\section{Introduction}

In recent years, optomechanical systems,
both in the optical and in the microwave regime, have
become one of the most prominent platforms for the investigation of
quantum mechanical phenomena. On the one hand, they have allowed
scientists to explore foundational aspects of quantum
theory~\cite{Marshall.2003,Aspelmeyer.2014}; on the other, they have
provided the test bed for future technological applications of quantum
mechanics~\cite{Barzanjeh.2022}. Prominent results in the field
include sideband~\cite{Teufel.2011owc} and feedback~\cite{Rossi.2018}
cooling to the ground state,
squeezing~\cite{Wollman.2015,Pirkkalainen.20150ene}, and
entanglement~\cite{Riedinger.2018,Ockeloen-Korppi.2018} of mechanical
resonators.  In the context of coupling between qubits and mechanical
resonators, the generation of quantum states of mechanical motion has
been recently realized in high-overtone bulk acoustic-wave resonators
(HBARs) where the generation of Fock~\cite{Chu.2018} and cat~\cite{Bild.2023} states were demonstrated.

In most of the examples mentioned above, the optomechanical system
consists of a mechanical resonator (e.g.~a nano-drum) whose position
is parametrically coupled to a photon cavity. One of the outstanding
goals in these systems has been the realization of the so-called
single-photon strong-coupling limit. In this regime, the parametric
coupling energy between a single photon and the mechanical mode
becomes comparable to the bare optical cavity linewidth and can
therefore significantly alter the dynamics of the system
~\cite{Heikkila.2014q2, Nation.2016,
  Romero-Sanchez.2018,Neumeier.2018,Neumeier.2018wjj,
  Settineri.2018ay, Kounalakis.2020,Liao.2020,Manninen.2022}. In
microwave setups, several proposals suggest that the addition of
nonlinear elements, in the form of Josephson junctions, can provide
the resource needed to reach the strong-coupling regime; realizations
along these lines include
charge~\cite{Heikkila.2014q2,Pirkkalainen.2015} and flux-mediated
optomechanical circuits~\cite{Bera.2021}.

Another intriguing aspect of these systems is
the possibility of realizing a quadratic coupling between the
mechanical motion and the optical field. Arguably, its most prominent
application is the detection of phonon Fock
states~\cite{Thompson.2008,
  Jayich.2008,Helmer.2009,Miao.2009,Nunnenkamp.2010}, even though
two-photon cooling and squeezing both of the mechanical and the
electromagnetic degrees of freedom have been
predicted~\cite{Nunnenkamp.2010} as well. In the optical frequencies range,
the quadratic coupling of an optical cavity with a mechanical mode has
been realized, e.g. in membrane-in-the-middle \cite{Thompson.2008}
and ultracold gases setups \cite{Purdy.2010}. In the microwave regime,
quadratic parametric coupling of a qubit to mechanical motion was
recently realized with drumhead mechanical resonators coupled to
superconducting circuits, exploiting the large mismatch of mechanical
and qubit resonant frequency ~\cite{Viennot.2018i3,Ma.2021}, where the
generation of (non-Gaussian) number-squeezed states was demonstrated.

\begin{figure*}[t]
  \includegraphics[width=0.95\textwidth]{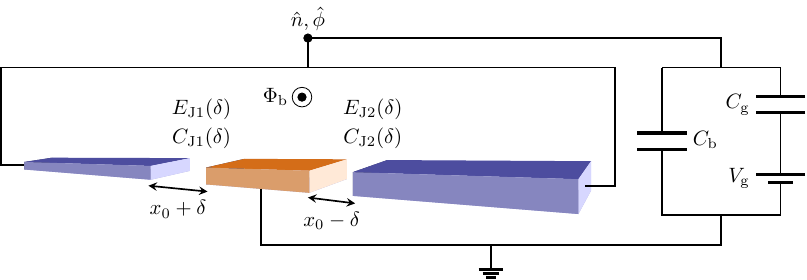}
  \caption{\label{fig:1}Cartoon picture of a (grounded) X2MON shuttle and a
    lumped-element description of a transmon-like setup, including the
    X2MON. Contacts between the shuttle (orange) and its terminals
    (blue) can be described by position-dependent 
    Josephson junctions.}
\end{figure*}

In this work, we extend the nonlinear circuit approaches mentioned
above, integrating a mechanical shuttling element into the design of the
superconducting circuit of Fig.~\ref{fig:1}. 
%Such electron shuttles have been demonstrated in detail in previous work, but have not 
%been realized as superconducting shuttles yet \cite{Scheible.2004,Koenig.2008,Kim.2012} 
%and only suggested being used from a theoretical point of view \cite{Gorelik.2001}. 
%In light of the above-mentioned connection of phonon states to the ongoing work in 
%superconducting qubits, this particular single electron shuttling mechanism certainly 
%offers a new %degree of freedom~\cite{Kim.2020}.
The shuttling element consists here of a portion of superconducting material that is free to 
perform mechanical oscillations between two (superconducting) electrodes. Analogous shuttling 
devices were realized experimentally in normal (i.e. non superconducting) circuits~\cite{Scheible.2004,Koenig.2008,Kim.2012},
demonstrating the ability of such devices to ``shuttle'' electrons along with the oscillatory 
mechanical motion. In addition, an analogous shuttling mechanism for Cooper pairs was theoretically 
investigated for superconducting circuits~\cite{Gorelik.2001}. In our work, we explore how the 
dynamical properties of a superconducting shuttling element can be recast in terms of a (nonlinear) 
optomechanical coupling between a superconducting circuit and a mechanical mode, for which 
we believe this particular charge shuttling mechanism certainly offers a new degree of 
freedom~\cite{Kim.2020}.

More specifically, we show how, in a lumped-element description, we are able to define a system 
constituted by a superconducting qubit exhibiting an intrinsic quadratic coupling to the mechanical
motion, in addition to a tunable linear one. The latter can be externally suppressed, leading to a 
dominant coupling that is quadratic in the mechanical degrees of freedom. At the same time, we
will show that the tunability of the linear coupling term allows for a coherent state exchange between 
the qubit and the mechanical resonator.

%

%xxxxxxxxxxxxxxxxxxxxxxxxxxxxxxxx
% DEVICE
%XXXXXXXXXXXXXXXXXXXXXXXXXXXXXXXX
\section{The device}

Our device is constituted by a 
superconducting shuttle, which is free to oscillate between two
terminals, as shown in Fig. 1.  
The terminals being gated to a voltage source $V_\mathrm{g}$
through a gating capacitance $C_\mathrm{g}$. The addition of a
shunting capacitance $C_\mathrm{b}$, meant to ensure protection from
charge fluctuations, defines a transmon qubit-like
device~\cite{Koch.2007} (the {X2MON}) exhibiting nontrivial properties
as a function of the mechanical shuttle dynamics.  The displacement of
the grounded island induces a shift in the Josephson and charging
energies of the two Josephson junctions (JJs), which translates into a
coupling between the superconducting circuit and the mechanical
motion. As anticipated, in our device, the coupling between the
mechanical degrees of freedom and a superconducting qubit can be
externally tuned between a \emph{linear} and a \emph{quadratic}
coupling, depending on the external magnetic flux through the loop
defined by the two JJs.  The setup we
propose here differs from the design of
Refs.~\cite{Viennot.2018i3,Ma.2021} inasmuch our realization, for
suitable values of the control parameter, is \textit{intrinsically}
quadratic in the mechanical displacement --i.e. not relying on the
relative value of the mechanical and qubit frequencies-- owing to the
symmetry of the design.

From a quantum-computational perspective, the coupling between a qubit
and a bosonic mode --represented here, as we will show, by a shuttling
mechanical element-- is an extremely promising candidate for quantum
error correction~\cite{Michael.2016,Sivak.2023}. Most importantly, the
tunability of the qubit/mechanical coupling of our setup allows for a
great flexibility in the choice of different protocols for
state-preparation and state-transfer between qubit and mechanics.
Further advantages offered by mechanical resonators compared to
microwave cavities as the bosonic mode are represented by their larger
coherence times (mechanical linewidths $\sim 1\, \mathrm{kHz}$~\cite{Koenig.2008}
vs. cavity linewidths $\sim 100\, \mathrm{kHz}$), the lack of
``cross-talk'' between the bosonic (mechanical) modes and, with specific
reference to shuttling mechanical elements, the scalability of such
platforms.

\subsection{Lumped-element model} 

Following Koch \textit{et al.}~\cite{Koch.2007}, 
we describe the Josephson junction energy as
the sum of a capacitive contribution $E_{\mathrm{C}}$ and the
Josephson energy $E_{\mathrm{J}}$.  In our device, the lower portion
of the circuit --the orange element in Fig.~\ref{fig:1}-- corresponds
to the actual island. As a consequence, the charging and Josephson
energies become dependent on the shuttle dynamics
$E_\mathrm{C\,1,2}=E_\mathrm{C\,1,2}(x_0\pm \delta)$,
$E_\mathrm{J\,1,2}=E_\mathrm{J\,1,2}(x_0\pm \delta)$. Owing to the
symmetry of the device, the upper (lower) sign corresponds to junction
1 (junction 2).

The total charging energy for the circuit can be written as
\begin{align}
  E_{\mathrm{C}}=\dfrac{e^{2}}{2C_\Sigma}
  \label{eq:1}  
\end{align}
with
$C_\Sigma=C_{\mathrm{J}1} +C_{\mathrm{J}2} +C_\mathrm{b}+C_\mathrm{g}$
where $C_{\mathrm{J}1}$ and $C_{\mathrm{J}2}$ are the capacitances of
the two Josephson junctions, $C_\mathrm{g}$ the gate capacitance and
$C_\mathrm{b}$ the shunting capacitance aimed at reducing the effects
of charge noise, like in a conventional transmon setup.

In the following, we will assume that the geometric capacitances
associated with the Josephson junctions $C_{\mathrm{J}1} $ and
$C_{\mathrm{J}2}$ can be modeled by two (equal, parallel plate)
capacitors
$C_{\mathrm{J}\,1,2} \doteq C_{\mathrm{J}}/\left(1 \pm \delta/x_0\right)$.
Furthermore, we assume an exponential dependence of $E_{\mathrm{J}}$ on the
electrodes' separation. This assumption can be justified through the
standard Ambegaokar-Baratoff formula \cite{Ambegaokar.1963,Martinis.2004} arguing that the normal-state resistance is given by
$R_\mathrm{N}=R_\mathrm{N0} \exp\left[x/\xi\right] $,
($x$ thickness of the JJ),
as a consequence of the exponential suppression of the tunneling
probability through a potential barrier, combined with the Landauer
formula \cite{Landauer.1957}, leading to
\begin{align}
  E_{\mathrm{J}\,1,2}=E_{\mathrm{J}} \exp \left[\mp \frac{\delta}{\xi} \right]
  \label{eq:2}
\end{align}
with $\xi=x_0/\log\left[\frac{\Delta R_K} {8 E_{\mathrm{J}} R_{\mathrm{N}0}}\right]$ 
and $E_{\mathrm{J}} = E_{\mathrm{J}1} (0) = E_{\mathrm{J}2} (0)$ for symmetrical JJs. 
Here $\Delta$ is the superconducting gap and 
$R_\mathrm{K} = 2 \pi \hbar / e^2$ the resistance quantum. 
For a typical $\mathrm{NbN}$ junction, we can assume $\Delta= 4500\,\mathrm{GHz}$, 
$E_{\mathrm{J}}= 20 \, \mathrm{GHz}$, 
$R_\mathrm{N0}= 50 \,\Omega$, and
$x_0 = 1\, \mathrm{nm}$, allowing us to estimate
$\xi \simeq 0.1 \, \mathrm{nm}$ ($\hbar=1$ throughout the manuscript).
Following Ref.~\cite{Koch.2007}, we can write the system's Hamiltonian as
\begin{align}
     H =& \, 4 E_{\mathrm{C}}(\delta) (n-n_\mathrm{g})^2 
        - E_\Sigma(\delta) \cos(\frac{\phi_\mathrm{b}}{2}) \cos(\phi) \nonumber\\
        &- E_\Delta(\delta) \sin(\frac{\phi_\mathrm{b}}{2}) \sin(\phi)+E\left(x_\mathrm{m},p_\mathrm{m}\right),
\label{eq:3}
\end{align}
where $n$ and $\phi$ are the excess number of charge carriers and
phase on the island, respectively, and
$ E_\Sigma(\delta) = E_{J1}(\delta) + E_{J2}(\delta)$,
$ E_\Delta(\delta) = E_{J1}(\delta) - E_{J2}(\delta)$.  For a
symmetric setup, we have
\begin{subequations}
  \begin{align}
    E_\Sigma(\delta) &= 2 E_{\mathrm{J}} \cosh\left(\frac{\delta}{\xi}\right) \label{eq:4}\\
    E_\Delta(\delta) &= 2 E_{\mathrm{J}} \sinh\left(\frac{\delta}{\xi}\right), \label{eq:5}
  \end{align}
\end{subequations}
where $E_\mathrm{J}$ is the Josephson energy associated with either
junction.  Furthermore, $n_\mathrm{g}$ is related to the external bias
voltage by $n_\mathrm{g} = -C_\mathrm{g} V_\mathrm{g}/2e$ and
the phase $\phi_{\rm b}$ is determined by the external flux bias
$\Phi_\mathrm{b}=\Phi_0/2\pi\, \phi_\mathrm{b}$ through the loop
defined by the two JJs ($\Phi_0=h/2e$). Here, the JJs are assumed to
be symmetrical, but the full calculation with general junctions is
presented in the Supplemental Material. Finally, the term
$E\left(x_\mathrm{m},p_\mathrm{m}\right)$ in the energy
  associated with the dynamics of the center of mass of the shuttle,
  which, in our analysis, we model as a simple harmonic oscillator.

Expanding the Hamiltonian given in Eq.~\eqref{eq:3} in powers of
$\hat{\phi}$ up to the fourth order and in powers of the mechanical
displacement up to the second order, focusing on the two lowest-charge
states we have (see e.g.~\cite{Girvin.2011}) 
\begin{equation}
\label{eq:6}
\begin{split}
  H =&\,  \frac{\omega_{\mathrm{q}}}{2} \, \sigma_\mathrm{z} + \omega_\mathrm{m} b^\dagger b \\
     & +  g_\mathrm{1} (b^\dagger + b) \sigma_\mathrm{x} 
       +  g_\mathrm{2} (b^\dagger + b)^2 \sigma_\mathrm{z}
\end{split}
\end{equation}
where $\sigma_\mathrm{z}=2 a^\dagger a-1$,
$\sigma_\mathrm{x} = a^\dagger + a$. The operators $a$ ($a^\dagger)$,
$b$ ($b^\dagger$) are the lowering (raising) operators associated with
the electrical and mechanical degrees of freedom, respectively.  All
coefficients in Eq.~\eqref{eq:6} ($ \omega_{\mathrm{q}}$,
$\omega_\mathrm{m}$, $g_\mathrm{1}$, $g_\mathrm{2}$) depend on the
external flux bias. The explicit dependence on $\Phi_\mathrm{b}$ is
given in the Supplemental Material.
As shown in Fig.~\ref{fig:2}, $g_\mathrm{2}$ is the dominating
interaction when the bias flux through the JJ loop is set to zero,
since $g_\mathrm{1}$ vanishes in this case.  However, $g_\mathrm{1}$
becomes the dominating term even for small deviations from the
$\Phi_\mathrm{b}=0$ condition. As we will show below, this ability to
control the type of interaction between the qubit and the mechanics
makes this system very flexible in terms of possible applications,
such as preparing the mechanical oscillator into a specific quantum
state. For $m \simeq 5\cdot 10^{-19}$ kg, $\omega_\mathrm{m}=1$ GHz,
and $\omega_\mathrm{q}\simeq 17$ GHz (all other parameters defined
above), we have that $x_{ZPF}/\xi \simeq 3\cdot 10^{-3}$,
$g_\mathrm{2}\simeq 40$ kHz.
 
\begin{figure}
  \includegraphics[width=0.45\textwidth]{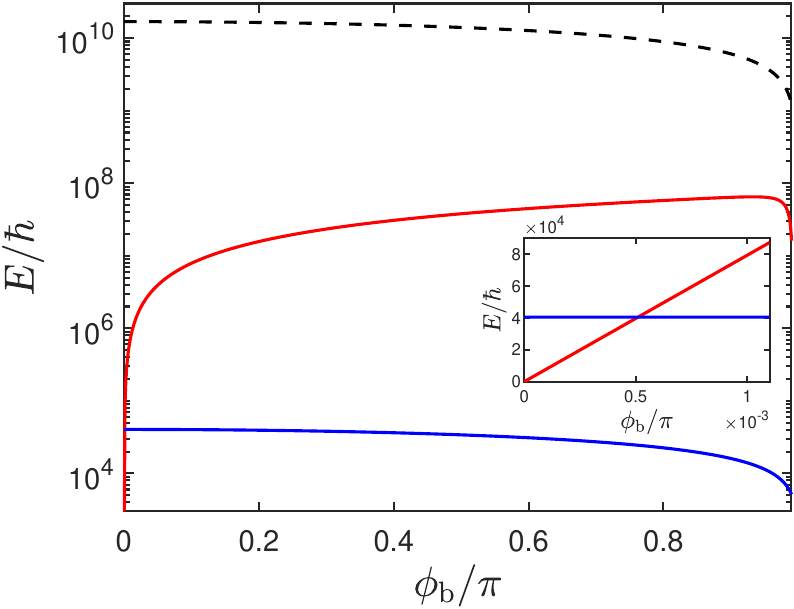}
  \caption{The strengths of $g_1$ (red) and
    $g_2$ (blue) couplings and the qubit frequency 
    $\omega_\mathrm{q}$ (dashed black) 
    as a function of the external flux $\phi_{\rm b}$. 
    The inset shows the crossover point near zero flux bias where
    $g_1$ becomes larger than $g_2$. Parameters
    $E_{\mathrm{J}} = 20\,$GHz, $E_{\mathrm{C}} = 1\,$GHz,
    $x_\mathrm{ZPF}/\xi = 3 \cdot 10^{-3}$.}
  \label{fig:2}
\end{figure}
Also, note that the Hamiltonian given in Eq.~\eqref{eq:6} is valid in
the case of equal JJs. If the JJs exhibit some degree of asymmetry, a
linear coupling to $\sigma_{\rm z}$ and a quadratic coupling to
$\sigma_{\rm x}$ appear, in addition to a small qubit rotation
(zeroth-order term in $\sigma_x$). While the quadratic coupling to
$\sigma_{\rm x}$ is negligible for all parameters, the linear coupling
between the displacement and $\sigma_z$ becomes comparable to the
quadratic one when the asymmetry of the Josephson energies is
$\sim \frac{x_{\rm ZPF}}{2\xi}$, which, for the parameters chosen
here, corresponds to
$\Delta E_{\rm J}=\left|E_{\rm J1}-E_{\rm J2}\right| \simeq 30\, {\rm MHz}$.
Furthermore, the zeroth-order term in $\sigma_x$ is negligible whenever
the asymmetry
$d_0=\left|E_{\rm J1}-E_{\rm J2}\right|/\left|E_{\rm J1}+E_{\rm J2}\right|$
is (much) less than unity. A discussion of the general case of
different JJs is presented in the Supplemental Material.
\begin{figure*}[ht]
  \includegraphics[width=0.95\textwidth]{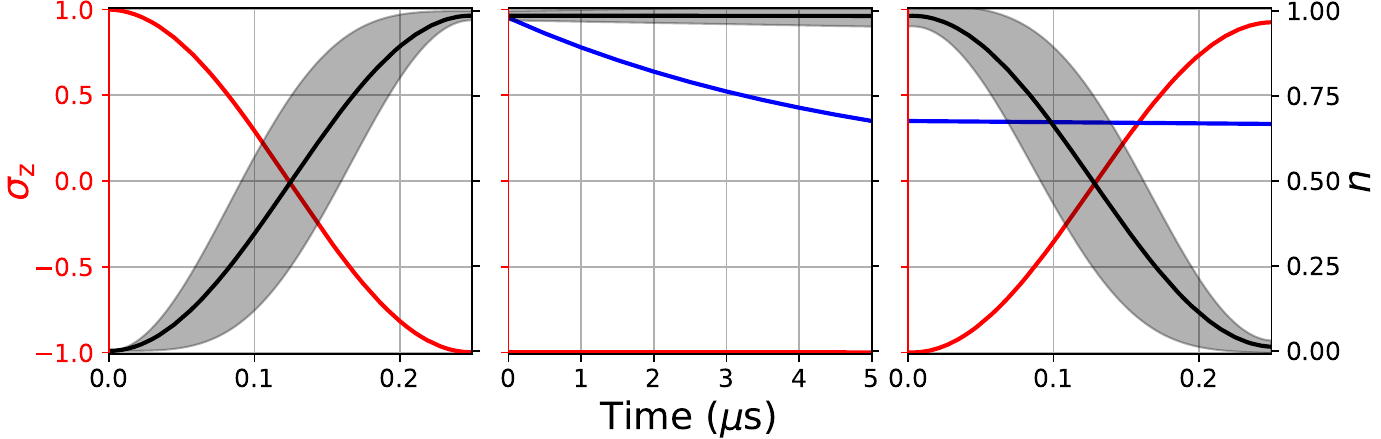}
  \caption{\label{fig:3} Demonstration of the state swapping
    protocol. \textit{Left panel:} the qubit state
    $\ket{\sigma_\mathrm{z}=1}$ (red line) is transferred to the
    mechanical oscillator Fock state $\ket{n=1}$ (black line/black
    shade). \textit{Central panel:} the linear coupling $g_\mathrm{1}$
    is suppressed. The free evolution of the state $\ket{n=1}$ is
    subject to the (small) intrinsic dissipation of the mechanical
    resonator. \textit{Right panel:} the state $\ket{n=1}$ is
    transferred back to the qubit. The fidelity of the swapping to the
    mechanical resonator and back, is compared with the free evolution
    of the $\ket{\sigma_\mathrm{z}=1}$ state under subject to pure
    qubit decoherence (blue line). 
    Parameters: $E_\mathrm{C}=1 \mathrm{\,GHz}$,
    $E_\mathrm{J}=20 \mathrm{\,GHz}$,
    $x_\mathrm{ZPF}/\xi = 3 \cdot 10^{-3}$, $\phi_\mathrm{b,0}=0.5$,
    mechanical angular frequency $\omega_\mathrm{m}=1$ GHz and
    damping rate $\gamma=1 \mathrm{\,kHz}$, qubit damping
    rate $\gamma_\sigma=100 \mathrm{\,kHz}$. In the simulations, we have
    assumed a finite temperature of $10$ mK, both for the qubit and
    for the mechanical bath, corresponding to a population of
    $n_{\sigma, {\rm th}}\simeq  2 \cdot 10^{-7} $ and
    $n_{{\rm m}, {\rm th}}\simeq  0.87$, respectively.
    The time axis is reset to zero for each step in the 
    protocol to better illustrate the different time scales.}
\end{figure*}

%xxxxxxxxxxxxxxxxxxxxxxxxxxxxxxxx
% PROTOCOLS
%XXXXXXXXXXXXXXXXXXXXXXXXXXXXXXXX
\section{State swapping protocol} 

A promising application for the
setup proposed here is bosonic error correction. On general grounds,
bosonic error correction provides key advantages over quantum error
correction schemes utilizing multiple physical qubits, inasmuch it
eliminates the overheads and the potential issues arising from
cross-talk of multiple physical qubits. More specifically, our
mechanical implementation has further advantages over a microwave
cavity: mechanical resonators offer better ringdown times than
microwave cavities, shuttling resonators offer a more compact design
and relatively straightforward scalability, and the qubit/mechanics
(linear) coupling can be externally tuned.

As a first step in this direction, we consider a state-swap protocol
that demonstrates the ability of this device to coherently transfer a
qubit state to a quantum state of the mechanical resonator.  Given the
relatively large frequency mismatch between the qubit and the mechanics
($\omega_\mathrm{q}/\omega_\mathrm{m} \simeq 20$), direct transitions between
mechanics and qubit are highly non-resonant. To induce such sideband
transitions, we therefore modulate the flux through the JJs loop at a
frequency $\bar{\omega}$, corresponding to a phase modulation given by 
$\phi_\mathrm{b}(t)=\phi_\mathrm{b,0} \cos\left(\bar{\omega} t \right)$
(flux-driven sideband transitions).  This technique is analogous to
the one employed in Ref.~\cite{Strand.2013} for the case of a transmon
qubit coupled to a microwave cavity. In the context of optomechanical
systems, a similar approach was also considered in \cite{Ma.2021}, where
the gate charge modulation was used to induce single-phonon
transitions in a mechanical resonator coupled to a qubit. In general,
these techniques go under the name of ac-dither techniques
\cite{Blais.2007}.

As a consequence of the phase modulation, the coefficients appearing
in the definition of the Hamiltonian $H$ in Eq.~\eqref{eq:6} become 
time dependent.
\begin{equation}
\label{eq:7}
\begin{split}
  H =&\,  \frac{\omega_{\mathrm{q}}(t)}{2} \, \sigma_\mathrm{z} + \omega_\mathrm{m}(t) b^\dagger b \\
     & +  g_\mathrm{1}(t) (b^\dagger + b) \sigma_\mathrm{x} 
       +  g_\mathrm{2}(t) (b^\dagger + b)^2 \sigma_\mathrm{z}.
\end{split}
\end{equation}
It is possible to show that, in a non-uniformly rotating frame for the
qubit and the mechanics, if the driving frequency is chosen in such a
way that $\bar{\omega}=\bar{\omega}_q-\omega_{\rm m}$, with
$\bar{\omega}_{\rm q}=\left. \omega_{\rm q}\right|_{\phi_{\rm b}=0}-\delta_{\rm q}$,
$\delta_{\rm q}=\phi_{b0}^2 \omega_{\rm p0}/16$
and $\omega_{\mathrm{p}0} = \sqrt{16 E_\mathrm{C} E_\mathrm{J} 
\cos (\phi_\mathrm{b}/2)}$ for symmetrical JJs, the
transformed Hamiltonian corresponds to a state-swapping Hamiltonian
\begin{align}
  \label{eq:8}
  H' = g_{\rm sw} \left( b^\dagger \sigma_- + b \sigma_+ \right)
\end{align}
where 
$g_{\rm sw}= \bar{g}_1 J_0\left(\delta_\mathrm{q}/2 \bar{\omega}\right)$
with $J_0(x)$ being the 0th-order Bessel function and 
$\bar{g}_1$ the linear approximation of $g_1(t)$ with respect to $\phi_\mathrm{b,0}$.

Given that $\phi_\mathrm{b,0}$ is externally tunable, after the
state-swapping transition described above, it is possible to
externally suppress the linear coupling between the qubit and the
mechanical resonator.  Having this in mind, one can consider using the
mechanical mode as a low-decoherence memory element, owing to the
combined effect of the on-demand suppression of lowest-order coupling
between the qubit and the mechanical resonator, and the intrinsically long
decoherence times of the mechanical element.

In Fig.~\ref{fig:3}, we depict a simple instance of the state-swap
protocol. Starting from the state
$\ket{\sigma_\mathrm{z}=1, n_\mathrm{m}=0}$, we first perform a state
swap between the qubit state and the mechanics (left), followed by the
``free'' evolution in the absence of qubit-mechanics coupling --
$\phi_b=0$, implying $g_\mathrm{1}=0$-- (center), transferring then
the excitation back to the qubit (right). We have compared this
state-swapping protocol with the free evolution of the
$\ket{\sigma_\mathrm{z}=1}$ state in the presence of the same
environment (qubit decay rate $\gamma_\sigma= 100$ kHz) as for the
state-swap protocol. The $\ket{\sigma_\mathrm{z}=1, n_\mathrm{m}=0}$
initial state can be prepared resorting to a preliminary cooling step
of the mechanical mode. We would like to point out that, while the
parameters chosen here push the boundaries of what is experimentally
realizable, with values of the $f \cdot Q $ product of the order of
$10^{14}$, record values of $f \cdot Q = O(10^{18})$ have been
reported for bulk acoustic wave
resonators~\cite{Galliou.2013,Kharel.20188in}.

%xxxxxxxxxxxxxxxxxxxxxxxxxxxxxxxx
% DISCUSSION
%XXXXXXXXXXXXXXXXXXXXXXXXXXXXXXXX
\section{Experimental outlook and performance}

From an experimental point of view, the realization of a shuttling island (orange box in Fig.~\ref{fig:1}) can be 
performed with some additional processing steps. However, fabrication of a superconducting 
island has turned out to be a challenge so far, since standard superconducting materials, 
such as Al, tend to oxidize through for small $50^3$nm$^3$ islands. Our latest work on that 
end shows that we can realize superconducting NbN-strips~\cite{Gonzalez.2023} 
with a $T_c =9$K, which avoids the aforementioned issue and is fully compatible with our 
processing techniques. Hence, the fabrication of superconducting islands in varying circuit 
combinations is possible now. One of the circuits to be implemented is shown in Fig. 1. 

Additionally, the ability to perform precise ac-dither protocols with the magnetic flux is 
required to successfully operate the device as a platform for useful quantum operations. These 
techniques are, in principle, possible with tuning either the gate charge or the magnetic 
flux~\cite{Blais.2007}, and different kinds of modulation schemes have been experimentally demonstrated 
to be feasible, e.g. Ref.\cite{Ma.2021} for charge-based and Refs.~\cite{Strand.2013,Valery.2022} 
for flux-based procedures.

Note that, in this work, we do not focus on parameter optimization for our 
device, for example, to maximize the couplings or to minimize the noise in the 
circuit, but instead we use fairly typical values for superconducting 
circuits. As an example, one could introduce a larger shunting capacitance to avoid 
charge noise effects thanks to the reduced charging energy, placing the 
device firmly in the $E_\mathrm{J} \gg E_\mathrm{C}$ transmon regime.

Even though we are not aiming to optimize the performance of our device here,
the accessible coupling strengths between the qubit and the mechanics
are promising when considering the proposed device as a candidate for 
practical quantum computation protocols. We obtain a quadratic coupling 
$g_2$ that is about an order of magnitude larger than presently achievable  
linewidths for shuttling mechanical 
resonators using the parameters in Fig.~\ref{fig:2} 
for a wide range of flux biasing, and
$g_2/\omega_\mathrm{m} \approx 4\times10^{-5}$  
which is comparable to different state-of-the-art implementations
of bosonic error correction schemes. For example, in a recent demonstration of 
quantum error correction beyond the break-even point using a bosonic (photonic) 
GKP code~\cite{Sivak.2023}, the corresponding ratio was $\sim 1\times10^{-5}$ 
with a superconducting cavity as the bosonic system. However, we do not predict reaching the ultrastrong
coupling limit as in Ref.~\cite{Ma.2021}, but in terms of the linear coupling $g_1$, 
we still can attain significantly larger values than the resonator linewidth
with $g_1/\omega_\mathrm{m} > 0.01$ already starting from a small flux bias.

We want to emphasize that the state swapping protocol explored in this work 
is not meant to be a comprehensive procedure for bosonic error correction, 
but instead a proof of concept that our device could be useful in such
applications. Indeed, it is possible to create arbitrary phonon number states 
with detailed control of the state swapping Hamiltonian~\cite{Law.1996} and, 
additionally, other interesting and useful operations can be 
realized with this interaction alone, such as the preparation of a mechanical 
cat states~\cite{Bild.2023}. On top of this linear interaction between the 
qubit and the mechanics, we also have access to another resource, namely
the quadratic coupling, that is largely unexplored in this work. 
%\noteJM{The following is a bit speculative and maybe could be removed.}
%One potential avenue to exploit the quadratic coupling would be to protect
%the reached Fock state after the linear coupling is turned off. This would
%be similar to the scheme in Ref.~\cite{Ma.2021}, except there the initial 
%squeezing of phonon number is also performed using the quadratic interaction.

%xxxxxxxxxxxxxxxxxxxxxxxxxxxxxxxx
% CONCLUSIONS
%XXXXXXXXXXXXXXXXXXXXXXXXXXXXXXXX
\section{Conclusions}

In our work, we introduce a novel device (the X2MON) which consists of a transmon 
coupled to a mechanical shuttle operating in the quantum regime. We discuss the nature of 
the coupling between the two. Furthermore, we demonstrate a state-swap protocol, which, 
owing to the properties of the shuttle, can be directly employed as a quantum memory. Our work paves the way for bosonic error correction with mechanical modes.

%e for controlling the interaction between a
%transmon qubit and a mechanical mode by using a moving Josephson
%junction shuttle. Being able to control the mechanical mode using a
%qubit paves the way towards initializing the mechanical resonator into
%a specific quantum state.

%xxxxxxxxxxxxxxxxxxxxxxxxxxxxxxxx
% ACKNOWLEDGMENTS
%XXXXXXXXXXXXXXXXXXXXXXXXXXXXXXXX
\begin{acknowledgments}
The numerical simulations were performed using the
\verb|QuantumOptics.jl| numerical framework~\cite{Kramer.2018}.  FM
and JM acknowledge financial support from the Research Council of
Norway (Grant No. 333937) through participation in the QuantERA
ERA-NET Cofund in Quantum Technologies. RHB acknowledges collaboration 
with Nexperia GmbH, Germany.
\end{acknowledgments}

\bibliography{references_X2monB.bib}

%apsrev4-2.bst 2019-01-14 (MD) hand-edited version of apsrev4-1.bst
%Control: key (0)
%Control: author (8) initials jnrlst
%Control: editor formatted (1) identically to author
%Control: production of article title (0) allowed
%Control: page (0) single
%Control: year (1) truncated
%Control: production of eprint (0) enabled
\providecommand{\noopsort}[1]{}\providecommand{\singleletter}[1]{#1}%
\begin{thebibliography}{50}%
\makeatletter
\providecommand \@ifxundefined [1]{%
 \@ifx{#1\undefined}
}%
\providecommand \@ifnum [1]{%
 \ifnum #1\expandafter \@firstoftwo
 \else \expandafter \@secondoftwo
 \fi
}%
\providecommand \@ifx [1]{%
 \ifx #1\expandafter \@firstoftwo
 \else \expandafter \@secondoftwo
 \fi
}%
\providecommand \natexlab [1]{#1}%
\providecommand \enquote  [1]{``#1''}%
\providecommand \bibnamefont  [1]{#1}%
\providecommand \bibfnamefont [1]{#1}%
\providecommand \citenamefont [1]{#1}%
\providecommand \href@noop [0]{\@secondoftwo}%
\providecommand \href [0]{\begingroup \@sanitize@url \@href}%
\providecommand \@href[1]{\@@startlink{#1}\@@href}%
\providecommand \@@href[1]{\endgroup#1\@@endlink}%
\providecommand \@sanitize@url [0]{\catcode `\\12\catcode `\$12\catcode
  `\&12\catcode `\#12\catcode `\^12\catcode `\_12\catcode `\%12\relax}%
\providecommand \@@startlink[1]{}%
\providecommand \@@endlink[0]{}%
\providecommand \url  [0]{\begingroup\@sanitize@url \@url }%
\providecommand \@url [1]{\endgroup\@href {#1}{\urlprefix }}%
\providecommand \urlprefix  [0]{URL }%
\providecommand \Eprint [0]{\href }%
\providecommand \doibase [0]{https://doi.org/}%
\providecommand \selectlanguage [0]{\@gobble}%
\providecommand \bibinfo  [0]{\@secondoftwo}%
\providecommand \bibfield  [0]{\@secondoftwo}%
\providecommand \translation [1]{[#1]}%
\providecommand \BibitemOpen [0]{}%
\providecommand \bibitemStop [0]{}%
\providecommand \bibitemNoStop [0]{.\EOS\space}%
\providecommand \EOS [0]{\spacefactor3000\relax}%
\providecommand \BibitemShut  [1]{\csname bibitem#1\endcsname}%
\let\auto@bib@innerbib\@empty
%</preamble>
\bibitem [{\citenamefont {Marshall}\ \emph {et~al.}(2003)\citenamefont
  {Marshall}, \citenamefont {Simon}, \citenamefont {Penrose},\ and\
  \citenamefont {Bouwmeester}}]{Marshall.2003}%
  \BibitemOpen
  \bibfield  {author} {\bibinfo {author} {\bibfnamefont {W.}~\bibnamefont
  {Marshall}}, \bibinfo {author} {\bibfnamefont {C.}~\bibnamefont {Simon}},
  \bibinfo {author} {\bibfnamefont {R.}~\bibnamefont {Penrose}},\ and\ \bibinfo
  {author} {\bibfnamefont {D.}~\bibnamefont {Bouwmeester}},\ }\bibfield
  {title} {\bibinfo {title} {{Towards Quantum Superpositions of a Mirror}},\
  }\href {https://doi.org/10.1103/physrevlett.91.130401} {\bibfield  {journal}
  {\bibinfo  {journal} {Physical Review Letters}\ }\textbf {\bibinfo {volume}
  {91}},\ \bibinfo {pages} {130401} (\bibinfo {year} {2003})}\BibitemShut
  {NoStop}%
\bibitem [{\citenamefont {Aspelmeyer}\ \emph {et~al.}(2014)\citenamefont
  {Aspelmeyer}, \citenamefont {Kippenberg},\ and\ \citenamefont
  {Marquardt}}]{Aspelmeyer.2014}%
  \BibitemOpen
  \bibfield  {author} {\bibinfo {author} {\bibfnamefont {M.}~\bibnamefont
  {Aspelmeyer}}, \bibinfo {author} {\bibfnamefont {T.~J.}\ \bibnamefont
  {Kippenberg}},\ and\ \bibinfo {author} {\bibfnamefont {F.}~\bibnamefont
  {Marquardt}},\ }\bibfield  {title} {\bibinfo {title} {{Cavity
  optomechanics}},\ }\href {https://doi.org/10.1103/revmodphys.86.1391}
  {\bibfield  {journal} {\bibinfo  {journal} {Reviews of Modern Physics}\
  }\textbf {\bibinfo {volume} {86}},\ \bibinfo {pages} {1391 } (\bibinfo {year}
  {2014})}\BibitemShut {NoStop}%
\bibitem [{\citenamefont {Barzanjeh}\ \emph {et~al.}(2022)\citenamefont
  {Barzanjeh}, \citenamefont {Xuereb}, \citenamefont {Gröblacher},
  \citenamefont {Paternostro}, \citenamefont {Regal},\ and\ \citenamefont
  {Weig}}]{Barzanjeh.2022}%
  \BibitemOpen
  \bibfield  {author} {\bibinfo {author} {\bibfnamefont {S.}~\bibnamefont
  {Barzanjeh}}, \bibinfo {author} {\bibfnamefont {A.}~\bibnamefont {Xuereb}},
  \bibinfo {author} {\bibfnamefont {S.}~\bibnamefont {Gröblacher}}, \bibinfo
  {author} {\bibfnamefont {M.}~\bibnamefont {Paternostro}}, \bibinfo {author}
  {\bibfnamefont {C.~A.}\ \bibnamefont {Regal}},\ and\ \bibinfo {author}
  {\bibfnamefont {E.~M.}\ \bibnamefont {Weig}},\ }\bibfield  {title} {\bibinfo
  {title} {Optomechanics for quantum technologies},\ }\href
  {https://doi.org/10.1038/s41567-021-01402-0} {\bibfield  {journal} {\bibinfo
  {journal} {Nature Physics}\ }\textbf {\bibinfo {volume} {18}},\ \bibinfo
  {pages} {15} (\bibinfo {year} {2022})}\BibitemShut {NoStop}%
\bibitem [{\citenamefont {Teufel}\ \emph {et~al.}(2011)\citenamefont {Teufel},
  \citenamefont {Donner}, \citenamefont {Li}, \citenamefont {Harlow},
  \citenamefont {Allman}, \citenamefont {Cicak}, \citenamefont {Sirois},
  \citenamefont {Whittaker}, \citenamefont {Lehnert},\ and\ \citenamefont
  {Simmonds}}]{Teufel.2011owc}%
  \BibitemOpen
  \bibfield  {author} {\bibinfo {author} {\bibfnamefont {J.~D.}\ \bibnamefont
  {Teufel}}, \bibinfo {author} {\bibfnamefont {T.}~\bibnamefont {Donner}},
  \bibinfo {author} {\bibfnamefont {D.}~\bibnamefont {Li}}, \bibinfo {author}
  {\bibfnamefont {J.~W.}\ \bibnamefont {Harlow}}, \bibinfo {author}
  {\bibfnamefont {M.~S.}\ \bibnamefont {Allman}}, \bibinfo {author}
  {\bibfnamefont {K.}~\bibnamefont {Cicak}}, \bibinfo {author} {\bibfnamefont
  {A.~J.}\ \bibnamefont {Sirois}}, \bibinfo {author} {\bibfnamefont {J.~D.}\
  \bibnamefont {Whittaker}}, \bibinfo {author} {\bibfnamefont {K.~W.}\
  \bibnamefont {Lehnert}},\ and\ \bibinfo {author} {\bibfnamefont {R.~W.}\
  \bibnamefont {Simmonds}},\ }\bibfield  {title} {\bibinfo {title} {{Sideband
  cooling of micromechanical motion to the quantum ground state}},\ }\href
  {https://doi.org/10.1038/nature10261} {\bibfield  {journal} {\bibinfo
  {journal} {Nature}\ }\textbf {\bibinfo {volume} {475}},\ \bibinfo {pages}
  {359 } (\bibinfo {year} {2011})}\BibitemShut {NoStop}%
\bibitem [{\citenamefont {Rossi}\ \emph {et~al.}(2018)\citenamefont {Rossi},
  \citenamefont {Mason}, \citenamefont {Chen}, \citenamefont {Tsaturyan},\ and\
  \citenamefont {Schliesser}}]{Rossi.2018}%
  \BibitemOpen
  \bibfield  {author} {\bibinfo {author} {\bibfnamefont {M.}~\bibnamefont
  {Rossi}}, \bibinfo {author} {\bibfnamefont {D.}~\bibnamefont {Mason}},
  \bibinfo {author} {\bibfnamefont {J.}~\bibnamefont {Chen}}, \bibinfo {author}
  {\bibfnamefont {Y.}~\bibnamefont {Tsaturyan}},\ and\ \bibinfo {author}
  {\bibfnamefont {A.}~\bibnamefont {Schliesser}},\ }\bibfield  {title}
  {\bibinfo {title} {{Measurement-based quantum control of mechanical
  motion}},\ }\href {https://doi.org/10.1038/s41586-018-0643-8} {\bibfield
  {journal} {\bibinfo  {journal} {Nature}\ }\textbf {\bibinfo {volume} {563}},\
  \bibinfo {pages} {53 } (\bibinfo {year} {2018})}\BibitemShut {NoStop}%
\bibitem [{\citenamefont {Wollman}\ \emph {et~al.}(2015)\citenamefont
  {Wollman}, \citenamefont {Lei}, \citenamefont {Weinstein}, \citenamefont
  {Suh}, \citenamefont {Kronwald}, \citenamefont {Marquardt}, \citenamefont
  {Clerk},\ and\ \citenamefont {Schwab}}]{Wollman.2015}%
  \BibitemOpen
  \bibfield  {author} {\bibinfo {author} {\bibfnamefont {E.~E.}\ \bibnamefont
  {Wollman}}, \bibinfo {author} {\bibfnamefont {C.~U.}\ \bibnamefont {Lei}},
  \bibinfo {author} {\bibfnamefont {A.~J.}\ \bibnamefont {Weinstein}}, \bibinfo
  {author} {\bibfnamefont {J.}~\bibnamefont {Suh}}, \bibinfo {author}
  {\bibfnamefont {A.}~\bibnamefont {Kronwald}}, \bibinfo {author}
  {\bibfnamefont {F.}~\bibnamefont {Marquardt}}, \bibinfo {author}
  {\bibfnamefont {A.~A.}\ \bibnamefont {Clerk}},\ and\ \bibinfo {author}
  {\bibfnamefont {K.~C.}\ \bibnamefont {Schwab}},\ }\bibfield  {title}
  {\bibinfo {title} {{Quantum squeezing of motion in a mechanical
  resonator.}},\ }\href {https://doi.org/10.1126/science.aac5138} {\bibfield
  {journal} {\bibinfo  {journal} {Science}\ }\textbf {\bibinfo {volume}
  {349}},\ \bibinfo {pages} {952 } (\bibinfo {year} {2015})}\BibitemShut
  {NoStop}%
\bibitem [{\citenamefont {Pirkkalainen}\ \emph
  {et~al.}(2015{\natexlab{a}})\citenamefont {Pirkkalainen}, \citenamefont
  {Damskägg}, \citenamefont {Brandt}, \citenamefont {Massel},\ and\
  \citenamefont {Sillanpää}}]{Pirkkalainen.20150ene}%
  \BibitemOpen
  \bibfield  {author} {\bibinfo {author} {\bibfnamefont {J.~M.}\ \bibnamefont
  {Pirkkalainen}}, \bibinfo {author} {\bibfnamefont {E.}~\bibnamefont
  {Damskägg}}, \bibinfo {author} {\bibfnamefont {M.}~\bibnamefont {Brandt}},
  \bibinfo {author} {\bibfnamefont {F.}~\bibnamefont {Massel}},\ and\ \bibinfo
  {author} {\bibfnamefont {M.~A.}\ \bibnamefont {Sillanpää}},\ }\bibfield
  {title} {\bibinfo {title} {{Squeezing of Quantum Noise of Motion in a
  Micromechanical Resonator}},\ }\href
  {https://doi.org/10.1103/physrevlett.115.243601} {\bibfield  {journal}
  {\bibinfo  {journal} {Physical Review Letters}\ }\textbf {\bibinfo {volume}
  {115}},\ \bibinfo {pages} {243601} (\bibinfo {year}
  {2015}{\natexlab{a}})}\BibitemShut {NoStop}%
\bibitem [{\citenamefont {Riedinger}\ \emph {et~al.}(2018)\citenamefont
  {Riedinger}, \citenamefont {Wallucks}, \citenamefont {Marinkovic},
  \citenamefont {Löschnauer}, \citenamefont {Aspelmeyer}, \citenamefont
  {Hong},\ and\ \citenamefont {Gröblacher}}]{Riedinger.2018}%
  \BibitemOpen
  \bibfield  {author} {\bibinfo {author} {\bibfnamefont {R.}~\bibnamefont
  {Riedinger}}, \bibinfo {author} {\bibfnamefont {A.}~\bibnamefont {Wallucks}},
  \bibinfo {author} {\bibfnamefont {I.}~\bibnamefont {Marinkovic}}, \bibinfo
  {author} {\bibfnamefont {C.}~\bibnamefont {Löschnauer}}, \bibinfo {author}
  {\bibfnamefont {M.}~\bibnamefont {Aspelmeyer}}, \bibinfo {author}
  {\bibfnamefont {S.}~\bibnamefont {Hong}},\ and\ \bibinfo {author}
  {\bibfnamefont {S.}~\bibnamefont {Gröblacher}},\ }\bibfield  {title}
  {\bibinfo {title} {{Remote quantum entanglement between two micromechanical
  oscillators}},\ }\href {https://doi.org/10.1038/s41586-018-0036-z} {\bibfield
   {journal} {\bibinfo  {journal} {Nature}\ }\textbf {\bibinfo {volume}
  {556}},\ \bibinfo {pages} {473 } (\bibinfo {year} {2018})}\BibitemShut
  {NoStop}%
\bibitem [{\citenamefont {Ockeloen-Korppi}\ \emph {et~al.}(2018)\citenamefont
  {Ockeloen-Korppi}, \citenamefont {Damskägg}, \citenamefont {Pirkkalainen},
  \citenamefont {Asjad}, \citenamefont {Clerk}, \citenamefont {Massel},
  \citenamefont {Woolley},\ and\ \citenamefont
  {Sillanpää}}]{Ockeloen-Korppi.2018}%
  \BibitemOpen
  \bibfield  {author} {\bibinfo {author} {\bibfnamefont {C.~F.}\ \bibnamefont
  {Ockeloen-Korppi}}, \bibinfo {author} {\bibfnamefont {E.}~\bibnamefont
  {Damskägg}}, \bibinfo {author} {\bibfnamefont {J.~M.}\ \bibnamefont
  {Pirkkalainen}}, \bibinfo {author} {\bibfnamefont {M.}~\bibnamefont {Asjad}},
  \bibinfo {author} {\bibfnamefont {A.~A.}\ \bibnamefont {Clerk}}, \bibinfo
  {author} {\bibfnamefont {F.}~\bibnamefont {Massel}}, \bibinfo {author}
  {\bibfnamefont {M.~J.}\ \bibnamefont {Woolley}},\ and\ \bibinfo {author}
  {\bibfnamefont {M.~A.}\ \bibnamefont {Sillanpää}},\ }\bibfield  {title}
  {\bibinfo {title} {{Stabilized entanglement of massive mechanical
  oscillators}},\ }\href {https://doi.org/10.1038/s41586-018-0038-x} {\bibfield
   {journal} {\bibinfo  {journal} {Nature}\ }\textbf {\bibinfo {volume}
  {556}},\ \bibinfo {pages} {478 } (\bibinfo {year} {2018})}\BibitemShut
  {NoStop}%
\bibitem [{\citenamefont {Chu}\ \emph {et~al.}(2018)\citenamefont {Chu},
  \citenamefont {Kharel}, \citenamefont {Yoon}, \citenamefont {Frunzio},
  \citenamefont {Rakich},\ and\ \citenamefont {Schoelkopf}}]{Chu.2018}%
  \BibitemOpen
  \bibfield  {author} {\bibinfo {author} {\bibfnamefont {Y.}~\bibnamefont
  {Chu}}, \bibinfo {author} {\bibfnamefont {P.}~\bibnamefont {Kharel}},
  \bibinfo {author} {\bibfnamefont {T.}~\bibnamefont {Yoon}}, \bibinfo {author}
  {\bibfnamefont {L.}~\bibnamefont {Frunzio}}, \bibinfo {author} {\bibfnamefont
  {P.~T.}\ \bibnamefont {Rakich}},\ and\ \bibinfo {author} {\bibfnamefont
  {R.~J.}\ \bibnamefont {Schoelkopf}},\ }\bibfield  {title} {\bibinfo {title}
  {{Creation and control of multi-phonon Fock states in a bulk acoustic-wave
  resonator}},\ }\href {https://doi.org/10.1038/s41586-018-0717-7} {\bibfield
  {journal} {\bibinfo  {journal} {Nature}\ }\textbf {\bibinfo {volume} {563}},\
  \bibinfo {pages} {666} (\bibinfo {year} {2018})}\BibitemShut {NoStop}%
\bibitem [{\citenamefont {Bild}\ \emph {et~al.}(2023)\citenamefont {Bild},
  \citenamefont {Fadel}, \citenamefont {Yang}, \citenamefont {Lüpke},
  \citenamefont {Martin}, \citenamefont {Bruno},\ and\ \citenamefont
  {Chu}}]{Bild.2023}%
  \BibitemOpen
  \bibfield  {author} {\bibinfo {author} {\bibfnamefont {M.}~\bibnamefont
  {Bild}}, \bibinfo {author} {\bibfnamefont {M.}~\bibnamefont {Fadel}},
  \bibinfo {author} {\bibfnamefont {Y.}~\bibnamefont {Yang}}, \bibinfo {author}
  {\bibfnamefont {U.~v.}\ \bibnamefont {Lüpke}}, \bibinfo {author}
  {\bibfnamefont {P.}~\bibnamefont {Martin}}, \bibinfo {author} {\bibfnamefont
  {A.}~\bibnamefont {Bruno}},\ and\ \bibinfo {author} {\bibfnamefont
  {Y.}~\bibnamefont {Chu}},\ }\bibfield  {title} {\bibinfo {title}
  {{Schrödinger cat states of a 16-microgram mechanical oscillator}},\ }\href
  {https://doi.org/10.1126/science.adf7553} {\bibfield  {journal} {\bibinfo
  {journal} {Science}\ }\textbf {\bibinfo {volume} {380}},\ \bibinfo {pages}
  {274} (\bibinfo {year} {2023})}\BibitemShut {NoStop}%
\bibitem [{\citenamefont {Heikkilä}\ \emph {et~al.}(2014)\citenamefont
  {Heikkilä}, \citenamefont {Massel}, \citenamefont {Tuorila}, \citenamefont
  {Khan},\ and\ \citenamefont {Sillanpää}}]{Heikkila.2014q2}%
  \BibitemOpen
  \bibfield  {author} {\bibinfo {author} {\bibfnamefont {T.~T.}\ \bibnamefont
  {Heikkilä}}, \bibinfo {author} {\bibfnamefont {F.}~\bibnamefont {Massel}},
  \bibinfo {author} {\bibfnamefont {J.}~\bibnamefont {Tuorila}}, \bibinfo
  {author} {\bibfnamefont {R.}~\bibnamefont {Khan}},\ and\ \bibinfo {author}
  {\bibfnamefont {M.~A.}\ \bibnamefont {Sillanpää}},\ }\bibfield  {title}
  {\bibinfo {title} {{Enhancing Optomechanical Coupling via the Josephson
  Effect}},\ }\href {https://doi.org/10.1103/physrevlett.112.203603} {\bibfield
   {journal} {\bibinfo  {journal} {Physical Review Letters}\ }\textbf {\bibinfo
  {volume} {112}},\ \bibinfo {pages} {203603} (\bibinfo {year}
  {2014})}\BibitemShut {NoStop}%
\bibitem [{\citenamefont {Nation}\ \emph {et~al.}(2016)\citenamefont {Nation},
  \citenamefont {Suh},\ and\ \citenamefont {Blencowe}}]{Nation.2016}%
  \BibitemOpen
  \bibfield  {author} {\bibinfo {author} {\bibfnamefont {P.~D.}\ \bibnamefont
  {Nation}}, \bibinfo {author} {\bibfnamefont {J.}~\bibnamefont {Suh}},\ and\
  \bibinfo {author} {\bibfnamefont {M.~P.}\ \bibnamefont {Blencowe}},\
  }\bibfield  {title} {\bibinfo {title} {{Ultrastrong optomechanics
  incorporating the dynamical Casimir effect}},\ }\href
  {https://doi.org/10.1103/physreva.93.022510} {\bibfield  {journal} {\bibinfo
  {journal} {Physical Review A}\ }\textbf {\bibinfo {volume} {93}},\ \bibinfo
  {pages} {022510} (\bibinfo {year} {2016})}\BibitemShut {NoStop}%
\bibitem [{\citenamefont {Romero-Sánchez}\ \emph {et~al.}(2018)\citenamefont
  {Romero-Sánchez}, \citenamefont {Bowen}, \citenamefont {Vanner},
  \citenamefont {Xia},\ and\ \citenamefont {Twamley}}]{Romero-Sanchez.2018}%
  \BibitemOpen
  \bibfield  {author} {\bibinfo {author} {\bibfnamefont {E.}~\bibnamefont
  {Romero-Sánchez}}, \bibinfo {author} {\bibfnamefont {W.~P.}\ \bibnamefont
  {Bowen}}, \bibinfo {author} {\bibfnamefont {M.~R.}\ \bibnamefont {Vanner}},
  \bibinfo {author} {\bibfnamefont {K.}~\bibnamefont {Xia}},\ and\ \bibinfo
  {author} {\bibfnamefont {J.}~\bibnamefont {Twamley}},\ }\bibfield  {title}
  {\bibinfo {title} {{Quantum magnetomechanics: Towards the ultrastrong
  coupling regime}},\ }\href {https://doi.org/10.1103/physrevb.97.024109}
  {\bibfield  {journal} {\bibinfo  {journal} {Physical Review B}\ }\textbf
  {\bibinfo {volume} {97}},\ \bibinfo {pages} {024109} (\bibinfo {year}
  {2018})}\BibitemShut {NoStop}%
\bibitem [{\citenamefont {Neumeier}\ \emph {et~al.}(2018)\citenamefont
  {Neumeier}, \citenamefont {Northup},\ and\ \citenamefont
  {Chang}}]{Neumeier.2018}%
  \BibitemOpen
  \bibfield  {author} {\bibinfo {author} {\bibfnamefont {L.}~\bibnamefont
  {Neumeier}}, \bibinfo {author} {\bibfnamefont {T.~E.}\ \bibnamefont
  {Northup}},\ and\ \bibinfo {author} {\bibfnamefont {D.~E.}\ \bibnamefont
  {Chang}},\ }\bibfield  {title} {\bibinfo {title} {{Reaching the
  optomechanical strong-coupling regime with a single atom in a cavity}},\
  }\href {https://doi.org/10.1103/physreva.97.063857} {\bibfield  {journal}
  {\bibinfo  {journal} {Physical Review A}\ }\textbf {\bibinfo {volume} {97}},\
  \bibinfo {pages} {063857} (\bibinfo {year} {2018})}\BibitemShut {NoStop}%
\bibitem [{\citenamefont {Neumeier}\ and\ \citenamefont
  {Chang}(2018)}]{Neumeier.2018wjj}%
  \BibitemOpen
  \bibfield  {author} {\bibinfo {author} {\bibfnamefont {L.}~\bibnamefont
  {Neumeier}}\ and\ \bibinfo {author} {\bibfnamefont {D.~E.}\ \bibnamefont
  {Chang}},\ }\bibfield  {title} {\bibinfo {title} {{Exploring unresolved
  sideband, optomechanical strong coupling using a single atom coupled to a
  cavity}},\ }\href {https://doi.org/10.1088/1367-2630/aad497} {\bibfield
  {journal} {\bibinfo  {journal} {New Journal of Physics}\ }\textbf {\bibinfo
  {volume} {20}},\ \bibinfo {pages} {083004} (\bibinfo {year}
  {2018})}\BibitemShut {NoStop}%
\bibitem [{\citenamefont {Settineri}\ \emph {et~al.}(2018)\citenamefont
  {Settineri}, \citenamefont {Macrí}, \citenamefont {Ridolfo}, \citenamefont
  {Stefano}, \citenamefont {Kockum}, \citenamefont {Nori},\ and\ \citenamefont
  {Savasta}}]{Settineri.2018ay}%
  \BibitemOpen
  \bibfield  {author} {\bibinfo {author} {\bibfnamefont {A.}~\bibnamefont
  {Settineri}}, \bibinfo {author} {\bibfnamefont {V.}~\bibnamefont {Macrí}},
  \bibinfo {author} {\bibfnamefont {A.}~\bibnamefont {Ridolfo}}, \bibinfo
  {author} {\bibfnamefont {O.~D.}\ \bibnamefont {Stefano}}, \bibinfo {author}
  {\bibfnamefont {A.~F.}\ \bibnamefont {Kockum}}, \bibinfo {author}
  {\bibfnamefont {F.}~\bibnamefont {Nori}},\ and\ \bibinfo {author}
  {\bibfnamefont {S.}~\bibnamefont {Savasta}},\ }\bibfield  {title} {\bibinfo
  {title} {{Dissipation and thermal noise in hybrid quantum systems in the
  ultrastrong-coupling regime}},\ }\href
  {https://doi.org/10.1103/physreva.98.053834} {\bibfield  {journal} {\bibinfo
  {journal} {Physical Review A}\ }\textbf {\bibinfo {volume} {98}},\ \bibinfo
  {pages} {053834} (\bibinfo {year} {2018})}\BibitemShut {NoStop}%
\bibitem [{\citenamefont {Kounalakis}\ \emph {et~al.}(2020)\citenamefont
  {Kounalakis}, \citenamefont {Blanter},\ and\ \citenamefont
  {Steele}}]{Kounalakis.2020}%
  \BibitemOpen
  \bibfield  {author} {\bibinfo {author} {\bibfnamefont {M.}~\bibnamefont
  {Kounalakis}}, \bibinfo {author} {\bibfnamefont {Y.~M.}\ \bibnamefont
  {Blanter}},\ and\ \bibinfo {author} {\bibfnamefont {G.~A.}\ \bibnamefont
  {Steele}},\ }\bibfield  {title} {\bibinfo {title} {{Flux-mediated
  optomechanics with a transmon qubit in the single-photon ultrastrong-coupling
  regime}},\ }\href {https://doi.org/10.1103/physrevresearch.2.023335}
  {\bibfield  {journal} {\bibinfo  {journal} {Physical Review Research}\
  }\textbf {\bibinfo {volume} {2}},\ \bibinfo {pages} {023335} (\bibinfo {year}
  {2020})}\BibitemShut {NoStop}%
\bibitem [{\citenamefont {Liao}\ \emph {et~al.}(2020)\citenamefont {Liao},
  \citenamefont {Huang}, \citenamefont {Tian}, \citenamefont {Kuang},\ and\
  \citenamefont {Sun}}]{Liao.2020}%
  \BibitemOpen
  \bibfield  {author} {\bibinfo {author} {\bibfnamefont {J.-Q.}\ \bibnamefont
  {Liao}}, \bibinfo {author} {\bibfnamefont {J.-F.}\ \bibnamefont {Huang}},
  \bibinfo {author} {\bibfnamefont {L.}~\bibnamefont {Tian}}, \bibinfo {author}
  {\bibfnamefont {L.-M.}\ \bibnamefont {Kuang}},\ and\ \bibinfo {author}
  {\bibfnamefont {C.-P.}\ \bibnamefont {Sun}},\ }\bibfield  {title} {\bibinfo
  {title} {{Generalized ultrastrong optomechanical-like coupling}},\ }\href
  {https://doi.org/10.1103/physreva.101.063802} {\bibfield  {journal} {\bibinfo
   {journal} {Physical Review A}\ }\textbf {\bibinfo {volume} {101}},\ \bibinfo
  {pages} {063802} (\bibinfo {year} {2020})}\BibitemShut {NoStop}%
\bibitem [{\citenamefont {Manninen}\ \emph {et~al.}(2022)\citenamefont
  {Manninen}, \citenamefont {Haque}, \citenamefont {Vitali},\ and\
  \citenamefont {Hakonen}}]{Manninen.2022}%
  \BibitemOpen
  \bibfield  {author} {\bibinfo {author} {\bibfnamefont {J.}~\bibnamefont
  {Manninen}}, \bibinfo {author} {\bibfnamefont {M.~T.}\ \bibnamefont {Haque}},
  \bibinfo {author} {\bibfnamefont {D.}~\bibnamefont {Vitali}},\ and\ \bibinfo
  {author} {\bibfnamefont {P.}~\bibnamefont {Hakonen}},\ }\bibfield  {title}
  {\bibinfo {title} {{Enhancement of the optomechanical coupling and Kerr
  nonlinearity using the Josephson capacitance of a Cooper-pair box}},\ }\href
  {https://doi.org/10.1103/physrevb.105.144508} {\bibfield  {journal} {\bibinfo
   {journal} {Physical Review B}\ }\textbf {\bibinfo {volume} {105}},\ \bibinfo
  {pages} {144508} (\bibinfo {year} {2022})}\BibitemShut {NoStop}%
\bibitem [{\citenamefont {Pirkkalainen}\ \emph
  {et~al.}(2015{\natexlab{b}})\citenamefont {Pirkkalainen}, \citenamefont
  {Cho}, \citenamefont {Massel}, \citenamefont {Tuorila}, \citenamefont
  {Heikkilä}, \citenamefont {Hakonen},\ and\ \citenamefont
  {Sillanpää}}]{Pirkkalainen.2015}%
  \BibitemOpen
  \bibfield  {author} {\bibinfo {author} {\bibfnamefont {J.~M.}\ \bibnamefont
  {Pirkkalainen}}, \bibinfo {author} {\bibfnamefont {S.~U.}\ \bibnamefont
  {Cho}}, \bibinfo {author} {\bibfnamefont {F.}~\bibnamefont {Massel}},
  \bibinfo {author} {\bibfnamefont {J.}~\bibnamefont {Tuorila}}, \bibinfo
  {author} {\bibfnamefont {T.~T.}\ \bibnamefont {Heikkilä}}, \bibinfo {author}
  {\bibfnamefont {P.~J.}\ \bibnamefont {Hakonen}},\ and\ \bibinfo {author}
  {\bibfnamefont {M.~A.}\ \bibnamefont {Sillanpää}},\ }\bibfield  {title}
  {\bibinfo {title} {{Cavity optomechanics mediated by a quantum two-level
  system}},\ }\href {https://doi.org/10.1038/ncomms7981} {\bibfield  {journal}
  {\bibinfo  {journal} {Nature communications}\ }\textbf {\bibinfo {volume}
  {6}},\ \bibinfo {pages} {6981} (\bibinfo {year}
  {2015}{\natexlab{b}})}\BibitemShut {NoStop}%
\bibitem [{\citenamefont {Bera}\ \emph {et~al.}(2021)\citenamefont {Bera},
  \citenamefont {Majumder}, \citenamefont {Sahu},\ and\ \citenamefont
  {Singh}}]{Bera.2021}%
  \BibitemOpen
  \bibfield  {author} {\bibinfo {author} {\bibfnamefont {T.}~\bibnamefont
  {Bera}}, \bibinfo {author} {\bibfnamefont {S.}~\bibnamefont {Majumder}},
  \bibinfo {author} {\bibfnamefont {S.~K.}\ \bibnamefont {Sahu}},\ and\
  \bibinfo {author} {\bibfnamefont {V.}~\bibnamefont {Singh}},\ }\bibfield
  {title} {\bibinfo {title} {Large flux-mediated coupling in hybrid
  electromechanical system with a transmon qubit},\ }\href
  {https://doi.org/10.1038/s42005-020-00514-y} {\bibfield  {journal} {\bibinfo
  {journal} {Communications Physics}\ }\textbf {\bibinfo {volume} {4}},\
  \bibinfo {pages} {12} (\bibinfo {year} {2021})}\BibitemShut {NoStop}%
\bibitem [{\citenamefont {Thompson}\ \emph {et~al.}(2008)\citenamefont
  {Thompson}, \citenamefont {Zwickl}, \citenamefont {Jayich}, \citenamefont
  {Marquardt}, \citenamefont {Girvin},\ and\ \citenamefont
  {Harris}}]{Thompson.2008}%
  \BibitemOpen
  \bibfield  {author} {\bibinfo {author} {\bibfnamefont {J.~D.}\ \bibnamefont
  {Thompson}}, \bibinfo {author} {\bibfnamefont {B.~M.}\ \bibnamefont
  {Zwickl}}, \bibinfo {author} {\bibfnamefont {A.~M.}\ \bibnamefont {Jayich}},
  \bibinfo {author} {\bibfnamefont {F.}~\bibnamefont {Marquardt}}, \bibinfo
  {author} {\bibfnamefont {S.~M.}\ \bibnamefont {Girvin}},\ and\ \bibinfo
  {author} {\bibfnamefont {J.~G.~E.}\ \bibnamefont {Harris}},\ }\bibfield
  {title} {\bibinfo {title} {Strong dispersive coupling of a high-finesse
  cavity to a micromechanical membrane},\ }\href
  {https://doi.org/10.1038/nature06715} {\bibfield  {journal} {\bibinfo
  {journal} {Nature}\ }\textbf {\bibinfo {volume} {452}},\ \bibinfo {pages}
  {72} (\bibinfo {year} {2008})}\BibitemShut {NoStop}%
\bibitem [{\citenamefont {Jayich}\ \emph {et~al.}(2008)\citenamefont {Jayich},
  \citenamefont {Sankey}, \citenamefont {Zwickl}, \citenamefont {Yang},
  \citenamefont {Thompson}, \citenamefont {Girvin}, \citenamefont {Clerk},
  \citenamefont {Marquardt},\ and\ \citenamefont {Harris}}]{Jayich.2008}%
  \BibitemOpen
  \bibfield  {author} {\bibinfo {author} {\bibfnamefont {A.~M.}\ \bibnamefont
  {Jayich}}, \bibinfo {author} {\bibfnamefont {J.~C.}\ \bibnamefont {Sankey}},
  \bibinfo {author} {\bibfnamefont {B.~M.}\ \bibnamefont {Zwickl}}, \bibinfo
  {author} {\bibfnamefont {C.}~\bibnamefont {Yang}}, \bibinfo {author}
  {\bibfnamefont {J.~D.}\ \bibnamefont {Thompson}}, \bibinfo {author}
  {\bibfnamefont {S.~M.}\ \bibnamefont {Girvin}}, \bibinfo {author}
  {\bibfnamefont {A.~A.}\ \bibnamefont {Clerk}}, \bibinfo {author}
  {\bibfnamefont {F.}~\bibnamefont {Marquardt}},\ and\ \bibinfo {author}
  {\bibfnamefont {J.~G.~E.}\ \bibnamefont {Harris}},\ }\bibfield  {title}
  {\bibinfo {title} {{Dispersive optomechanics: A membrane inside a cavity}},\
  }\href {https://doi.org/10.1088/1367-2630/10/9/095008} {\bibfield  {journal}
  {\bibinfo  {journal} {New Journal Of Physics}\ }\textbf {\bibinfo {volume}
  {10}},\ \bibinfo {pages} {095008} (\bibinfo {year} {2008})}\BibitemShut
  {NoStop}%
\bibitem [{\citenamefont {Helmer}\ \emph {et~al.}(2009)\citenamefont {Helmer},
  \citenamefont {Mariantoni}, \citenamefont {Solano},\ and\ \citenamefont
  {Marquardt}}]{Helmer.2009}%
  \BibitemOpen
  \bibfield  {author} {\bibinfo {author} {\bibfnamefont {F.}~\bibnamefont
  {Helmer}}, \bibinfo {author} {\bibfnamefont {M.}~\bibnamefont {Mariantoni}},
  \bibinfo {author} {\bibfnamefont {E.}~\bibnamefont {Solano}},\ and\ \bibinfo
  {author} {\bibfnamefont {F.}~\bibnamefont {Marquardt}},\ }\bibfield  {title}
  {\bibinfo {title} {{Quantum nondemolition photon detection in circuit QED and
  the quantum Zeno effect}},\ }\href
  {https://doi.org/10.1103/physreva.79.052115} {\bibfield  {journal} {\bibinfo
  {journal} {Physical Review A}\ }\textbf {\bibinfo {volume} {79}},\ \bibinfo
  {pages} {052115} (\bibinfo {year} {2009})}\BibitemShut {NoStop}%
\bibitem [{\citenamefont {Miao}\ \emph {et~al.}(2009)\citenamefont {Miao},
  \citenamefont {Danilishin}, \citenamefont {Corbitt},\ and\ \citenamefont
  {Chen}}]{Miao.2009}%
  \BibitemOpen
  \bibfield  {author} {\bibinfo {author} {\bibfnamefont {H.}~\bibnamefont
  {Miao}}, \bibinfo {author} {\bibfnamefont {S.}~\bibnamefont {Danilishin}},
  \bibinfo {author} {\bibfnamefont {T.}~\bibnamefont {Corbitt}},\ and\ \bibinfo
  {author} {\bibfnamefont {Y.}~\bibnamefont {Chen}},\ }\bibfield  {title}
  {\bibinfo {title} {{Standard Quantum Limit for Probing Mechanical Energy
  Quantization}},\ }\href {https://doi.org/10.1103/physrevlett.103.100402}
  {\bibfield  {journal} {\bibinfo  {journal} {Physical Review Letters}\
  }\textbf {\bibinfo {volume} {103}},\ \bibinfo {pages} {100402} (\bibinfo
  {year} {2009})}\BibitemShut {NoStop}%
\bibitem [{\citenamefont {Nunnenkamp}\ \emph {et~al.}(2010)\citenamefont
  {Nunnenkamp}, \citenamefont {Børkje}, \citenamefont {Harris},\ and\
  \citenamefont {Girvin}}]{Nunnenkamp.2010}%
  \BibitemOpen
  \bibfield  {author} {\bibinfo {author} {\bibfnamefont {A.}~\bibnamefont
  {Nunnenkamp}}, \bibinfo {author} {\bibfnamefont {K.}~\bibnamefont {Børkje}},
  \bibinfo {author} {\bibfnamefont {J.~G.~E.}\ \bibnamefont {Harris}},\ and\
  \bibinfo {author} {\bibfnamefont {S.~M.}\ \bibnamefont {Girvin}},\ }\bibfield
   {title} {\bibinfo {title} {{Cooling and squeezing via quadratic
  optomechanical coupling}},\ }\href
  {https://doi.org/10.1103/physreva.82.021806} {\bibfield  {journal} {\bibinfo
  {journal} {Physical Review A}\ }\textbf {\bibinfo {volume} {82}},\ \bibinfo
  {pages} {021806} (\bibinfo {year} {2010})}\BibitemShut {NoStop}%
\bibitem [{\citenamefont {Purdy}\ \emph {et~al.}(2010)\citenamefont {Purdy},
  \citenamefont {Brooks}, \citenamefont {Botter}, \citenamefont {Brahms},
  \citenamefont {Ma},\ and\ \citenamefont {Stamper-Kurn}}]{Purdy.2010}%
  \BibitemOpen
  \bibfield  {author} {\bibinfo {author} {\bibfnamefont {T.~P.}\ \bibnamefont
  {Purdy}}, \bibinfo {author} {\bibfnamefont {D.~W.~C.}\ \bibnamefont
  {Brooks}}, \bibinfo {author} {\bibfnamefont {T.}~\bibnamefont {Botter}},
  \bibinfo {author} {\bibfnamefont {N.}~\bibnamefont {Brahms}}, \bibinfo
  {author} {\bibfnamefont {Z.-Y.}\ \bibnamefont {Ma}},\ and\ \bibinfo {author}
  {\bibfnamefont {D.~M.}\ \bibnamefont {Stamper-Kurn}},\ }\bibfield  {title}
  {\bibinfo {title} {Tunable cavity optomechanics with ultracold atoms},\
  }\href {https://doi.org/10.1103/PhysRevLett.105.133602} {\bibfield  {journal}
  {\bibinfo  {journal} {Physical Review Letters}\ }\textbf {\bibinfo {volume}
  {105}},\ \bibinfo {pages} {133602} (\bibinfo {year} {2010})}\BibitemShut
  {NoStop}%
\bibitem [{\citenamefont {Viennot}\ \emph {et~al.}(2018)\citenamefont
  {Viennot}, \citenamefont {Ma},\ and\ \citenamefont
  {Lehnert}}]{Viennot.2018i3}%
  \BibitemOpen
  \bibfield  {author} {\bibinfo {author} {\bibfnamefont {J.~J.}\ \bibnamefont
  {Viennot}}, \bibinfo {author} {\bibfnamefont {X.}~\bibnamefont {Ma}},\ and\
  \bibinfo {author} {\bibfnamefont {K.~W.}\ \bibnamefont {Lehnert}},\
  }\bibfield  {title} {\bibinfo {title} {{Phonon-Number-Sensitive
  Electromechanics.}},\ }\href {https://doi.org/10.1103/physrevlett.121.183601}
  {\bibfield  {journal} {\bibinfo  {journal} {Physical Review Letters}\
  }\textbf {\bibinfo {volume} {121}},\ \bibinfo {pages} {183601} (\bibinfo
  {year} {2018})}\BibitemShut {NoStop}%
\bibitem [{\citenamefont {Ma}\ \emph {et~al.}(2021)\citenamefont {Ma},
  \citenamefont {Viennot}, \citenamefont {Kotler}, \citenamefont {Teufel},\
  and\ \citenamefont {Lehnert}}]{Ma.2021}%
  \BibitemOpen
  \bibfield  {author} {\bibinfo {author} {\bibfnamefont {X.}~\bibnamefont
  {Ma}}, \bibinfo {author} {\bibfnamefont {J.~J.}\ \bibnamefont {Viennot}},
  \bibinfo {author} {\bibfnamefont {S.}~\bibnamefont {Kotler}}, \bibinfo
  {author} {\bibfnamefont {J.~D.}\ \bibnamefont {Teufel}},\ and\ \bibinfo
  {author} {\bibfnamefont {K.~W.}\ \bibnamefont {Lehnert}},\ }\bibfield
  {title} {\bibinfo {title} {{Non-classical energy squeezing of a macroscopic
  mechanical oscillator}},\ }\href {https://doi.org/10.1038/s41567-020-01102-1}
  {\bibfield  {journal} {\bibinfo  {journal} {Nature Physics}\ }\textbf
  {\bibinfo {volume} {17}},\ \bibinfo {pages} {322} (\bibinfo {year}
  {2021})}\BibitemShut {NoStop}%
\bibitem [{\citenamefont {Scheible}\ and\ \citenamefont
  {Blick}(2004)}]{Scheible.2004}%
  \BibitemOpen
  \bibfield  {author} {\bibinfo {author} {\bibfnamefont {D.~V.}\ \bibnamefont
  {Scheible}}\ and\ \bibinfo {author} {\bibfnamefont {R.~H.}\ \bibnamefont
  {Blick}},\ }\bibfield  {title} {\bibinfo {title} {{Silicon nanopillars for
  mechanical single-electron transport}},\ }\href
  {https://doi.org/10.1063/1.1759371} {\bibfield  {journal} {\bibinfo
  {journal} {Applied Physics Letters}\ }\textbf {\bibinfo {volume} {84}},\
  \bibinfo {pages} {4632} (\bibinfo {year} {2004})}\BibitemShut {NoStop}%
\bibitem [{\citenamefont {Koenig}\ \emph {et~al.}(2008)\citenamefont {Koenig},
  \citenamefont {Weig},\ and\ \citenamefont {Kotthaus}}]{Koenig.2008}%
  \BibitemOpen
  \bibfield  {author} {\bibinfo {author} {\bibfnamefont {D.~R.}\ \bibnamefont
  {Koenig}}, \bibinfo {author} {\bibfnamefont {E.~M.}\ \bibnamefont {Weig}},\
  and\ \bibinfo {author} {\bibfnamefont {J.~P.}\ \bibnamefont {Kotthaus}},\
  }\bibfield  {title} {\bibinfo {title} {Ultrasonically driven nanomechanical
  single-electron shuttle},\ }\href {https://doi.org/10.1038/nnano.2008.178}
  {\bibfield  {journal} {\bibinfo  {journal} {Nature Nanotechnology}\ }\textbf
  {\bibinfo {volume} {3}},\ \bibinfo {pages} {482} (\bibinfo {year}
  {2008})}\BibitemShut {NoStop}%
\bibitem [{\citenamefont {Kim}\ \emph {et~al.}(2012)\citenamefont {Kim},
  \citenamefont {Prada},\ and\ \citenamefont {Blick}}]{Kim.2012}%
  \BibitemOpen
  \bibfield  {author} {\bibinfo {author} {\bibfnamefont {C.}~\bibnamefont
  {Kim}}, \bibinfo {author} {\bibfnamefont {M.}~\bibnamefont {Prada}},\ and\
  \bibinfo {author} {\bibfnamefont {R.~H.}\ \bibnamefont {Blick}},\ }\bibfield
  {title} {\bibinfo {title} {Coulomb blockade in a coupled nanomechanical
  electron shuttle},\ }\href {https://doi.org/10.1021/nn204103m} {\bibfield
  {journal} {\bibinfo  {journal} {ACS Nano}\ }\textbf {\bibinfo {volume} {6}},\
  \bibinfo {pages} {651} (\bibinfo {year} {2012})}\BibitemShut {NoStop}%
\bibitem [{\citenamefont {Gorelik}\ \emph {et~al.}(2001)\citenamefont
  {Gorelik}, \citenamefont {Isacsson}, \citenamefont {Galperin}, \citenamefont
  {Shekhter},\ and\ \citenamefont {Jonson}}]{Gorelik.2001}%
  \BibitemOpen
  \bibfield  {author} {\bibinfo {author} {\bibfnamefont {L.~Y.}\ \bibnamefont
  {Gorelik}}, \bibinfo {author} {\bibfnamefont {A.}~\bibnamefont {Isacsson}},
  \bibinfo {author} {\bibfnamefont {Y.~M.}\ \bibnamefont {Galperin}}, \bibinfo
  {author} {\bibfnamefont {R.~I.}\ \bibnamefont {Shekhter}},\ and\ \bibinfo
  {author} {\bibfnamefont {M.}~\bibnamefont {Jonson}},\ }\bibfield  {title}
  {\bibinfo {title} {{Coherent transfer of Cooper pairs by a movable grain}},\
  }\href {https://doi.org/10.1038/35078027} {\bibfield  {journal} {\bibinfo
  {journal} {Nature}\ }\textbf {\bibinfo {volume} {411}},\ \bibinfo {pages}
  {454} (\bibinfo {year} {2001})}\BibitemShut {NoStop}%
\bibitem [{\citenamefont {Kim}\ \emph {et~al.}(2020)\citenamefont {Kim},
  \citenamefont {Marsland},\ and\ \citenamefont {Blick}}]{Kim.2020}%
  \BibitemOpen
  \bibfield  {author} {\bibinfo {author} {\bibfnamefont {C.}~\bibnamefont
  {Kim}}, \bibinfo {author} {\bibfnamefont {R.}~\bibnamefont {Marsland}},\ and\
  \bibinfo {author} {\bibfnamefont {R.~H.}\ \bibnamefont {Blick}},\ }\bibfield
  {title} {\bibinfo {title} {The nanomechanical bit},\ }\href
  {https://doi.org/https://doi.org/10.1002/smll.202001580} {\bibfield
  {journal} {\bibinfo  {journal} {Small}\ }\textbf {\bibinfo {volume} {16}},\
  \bibinfo {pages} {2001580} (\bibinfo {year} {2020})}\BibitemShut {NoStop}%
\bibitem [{\citenamefont {Koch}\ \emph {et~al.}(2007)\citenamefont {Koch},
  \citenamefont {Yu}, \citenamefont {Gambetta}, \citenamefont {Houck},
  \citenamefont {Schuster}, \citenamefont {Majer}, \citenamefont {Blais},
  \citenamefont {Devoret}, \citenamefont {Girvin},\ and\ \citenamefont
  {Schoelkopf}}]{Koch.2007}%
  \BibitemOpen
  \bibfield  {author} {\bibinfo {author} {\bibfnamefont {J.}~\bibnamefont
  {Koch}}, \bibinfo {author} {\bibfnamefont {T.~M.}\ \bibnamefont {Yu}},
  \bibinfo {author} {\bibfnamefont {J.}~\bibnamefont {Gambetta}}, \bibinfo
  {author} {\bibfnamefont {A.~A.}\ \bibnamefont {Houck}}, \bibinfo {author}
  {\bibfnamefont {D.~I.}\ \bibnamefont {Schuster}}, \bibinfo {author}
  {\bibfnamefont {J.}~\bibnamefont {Majer}}, \bibinfo {author} {\bibfnamefont
  {A.}~\bibnamefont {Blais}}, \bibinfo {author} {\bibfnamefont {M.~H.}\
  \bibnamefont {Devoret}}, \bibinfo {author} {\bibfnamefont {S.~M.}\
  \bibnamefont {Girvin}},\ and\ \bibinfo {author} {\bibfnamefont {R.~J.}\
  \bibnamefont {Schoelkopf}},\ }\bibfield  {title} {\bibinfo {title}
  {{Charge-insensitive qubit design derived from the Cooper pair box}},\ }\href
  {https://doi.org/10.1103/physreva.76.042319} {\bibfield  {journal} {\bibinfo
  {journal} {Physical Review A}\ }\textbf {\bibinfo {volume} {76}},\ \bibinfo
  {pages} {042319} (\bibinfo {year} {2007})}\BibitemShut {NoStop}%
\bibitem [{\citenamefont {Michael}\ \emph {et~al.}(2016)\citenamefont
  {Michael}, \citenamefont {Silveri}, \citenamefont {Brierley}, \citenamefont
  {Albert}, \citenamefont {Salmilehto}, \citenamefont {Jiang},\ and\
  \citenamefont {Girvin}}]{Michael.2016}%
  \BibitemOpen
  \bibfield  {author} {\bibinfo {author} {\bibfnamefont {M.~H.}\ \bibnamefont
  {Michael}}, \bibinfo {author} {\bibfnamefont {M.}~\bibnamefont {Silveri}},
  \bibinfo {author} {\bibfnamefont {R.~T.}\ \bibnamefont {Brierley}}, \bibinfo
  {author} {\bibfnamefont {V.~V.}\ \bibnamefont {Albert}}, \bibinfo {author}
  {\bibfnamefont {J.}~\bibnamefont {Salmilehto}}, \bibinfo {author}
  {\bibfnamefont {L.}~\bibnamefont {Jiang}},\ and\ \bibinfo {author}
  {\bibfnamefont {S.~M.}\ \bibnamefont {Girvin}},\ }\bibfield  {title}
  {\bibinfo {title} {New class of quantum error-correcting codes for a bosonic
  mode},\ }\href {https://doi.org/10.1103/PhysRevX.6.031006} {\bibfield
  {journal} {\bibinfo  {journal} {Physical Review X}\ }\textbf {\bibinfo
  {volume} {6}},\ \bibinfo {pages} {031006} (\bibinfo {year}
  {2016})}\BibitemShut {NoStop}%
\bibitem [{\citenamefont {Sivak}\ \emph {et~al.}(2023)\citenamefont {Sivak},
  \citenamefont {Eickbusch}, \citenamefont {Royer}, \citenamefont {Singh},
  \citenamefont {Tsioutsios}, \citenamefont {Ganjam}, \citenamefont {Miano},
  \citenamefont {Brock}, \citenamefont {Ding}, \citenamefont {Frunzio},
  \citenamefont {Girvin}, \citenamefont {Schoelkopf},\ and\ \citenamefont
  {Devoret}}]{Sivak.2023}%
  \BibitemOpen
  \bibfield  {author} {\bibinfo {author} {\bibfnamefont {V.~V.}\ \bibnamefont
  {Sivak}}, \bibinfo {author} {\bibfnamefont {A.}~\bibnamefont {Eickbusch}},
  \bibinfo {author} {\bibfnamefont {B.}~\bibnamefont {Royer}}, \bibinfo
  {author} {\bibfnamefont {S.}~\bibnamefont {Singh}}, \bibinfo {author}
  {\bibfnamefont {I.}~\bibnamefont {Tsioutsios}}, \bibinfo {author}
  {\bibfnamefont {S.}~\bibnamefont {Ganjam}}, \bibinfo {author} {\bibfnamefont
  {A.}~\bibnamefont {Miano}}, \bibinfo {author} {\bibfnamefont {B.~L.}\
  \bibnamefont {Brock}}, \bibinfo {author} {\bibfnamefont {A.~Z.}\ \bibnamefont
  {Ding}}, \bibinfo {author} {\bibfnamefont {L.}~\bibnamefont {Frunzio}},
  \bibinfo {author} {\bibfnamefont {S.~M.}\ \bibnamefont {Girvin}}, \bibinfo
  {author} {\bibfnamefont {R.~J.}\ \bibnamefont {Schoelkopf}},\ and\ \bibinfo
  {author} {\bibfnamefont {M.~H.}\ \bibnamefont {Devoret}},\ }\bibfield
  {title} {\bibinfo {title} {Real-time quantum error correction beyond
  break-even},\ }\href {https://doi.org/10.1038/s41586-023-05782-6} {\bibfield
  {journal} {\bibinfo  {journal} {Nature}\ }\textbf {\bibinfo {volume} {616}},\
  \bibinfo {pages} {50} (\bibinfo {year} {2023})}\BibitemShut {NoStop}%
\bibitem [{\citenamefont {Ambegaokar}\ and\ \citenamefont
  {Baratoff}(1963)}]{Ambegaokar.1963}%
  \BibitemOpen
  \bibfield  {author} {\bibinfo {author} {\bibfnamefont {V.}~\bibnamefont
  {Ambegaokar}}\ and\ \bibinfo {author} {\bibfnamefont {A.}~\bibnamefont
  {Baratoff}},\ }\bibfield  {title} {\bibinfo {title} {{Tunneling Between
  Superconductors}},\ }\href {https://doi.org/10.1103/physrevlett.10.486}
  {\bibfield  {journal} {\bibinfo  {journal} {Physical Review Letters}\
  }\textbf {\bibinfo {volume} {10}},\ \bibinfo {pages} {486} (\bibinfo {year}
  {1963})}\BibitemShut {NoStop}%
\bibitem [{\citenamefont {Martinis}(2004)}]{Martinis.2004}%
  \BibitemOpen
  \bibfield  {author} {\bibinfo {author} {\bibfnamefont {J.~M.}\ \bibnamefont
  {Martinis}},\ }\bibfield  {title} {\bibinfo {title} {{Course 13
  Superconducting qubits and the physics of Josephson junctions}},\ }\href
  {https://doi.org/10.1016/s0924-8099(03)80037-9} {\bibfield  {journal}
  {\bibinfo  {journal} {Les Houches}\ }\textbf {\bibinfo {volume} {79}},\
  \bibinfo {pages} {487} (\bibinfo {year} {2004})}\BibitemShut {NoStop}%
\bibitem [{\citenamefont {Landauer}(1957)}]{Landauer.1957}%
  \BibitemOpen
  \bibfield  {author} {\bibinfo {author} {\bibfnamefont {R.}~\bibnamefont
  {Landauer}},\ }\bibfield  {title} {\bibinfo {title} {{Spatial Variation of
  Currents and Fields Due to Localized Scatterers in Metallic Conduction}},\
  }\href {https://doi.org/10.1147/rd.13.0223} {\bibfield  {journal} {\bibinfo
  {journal} {IBM Journal of Research and Development}\ }\textbf {\bibinfo
  {volume} {1}},\ \bibinfo {pages} {223} (\bibinfo {year} {1957})}\BibitemShut
  {NoStop}%
\bibitem [{\citenamefont {Girvin}(2011)}]{Girvin.2011}%
  \BibitemOpen
  \bibfield  {author} {\bibinfo {author} {\bibfnamefont {S.~M.}\ \bibnamefont
  {Girvin}},\ }\bibfield  {title} {\bibinfo {title} {{Superconducting qubits
  and circuits: Artificial atoms coupled to microwave photons}},\ }\href@noop
  {} {\bibfield  {journal} {\bibinfo  {journal} {École d’été Les Houches}\
  } (\bibinfo {year} {2011})}\BibitemShut {NoStop}%
\bibitem [{\citenamefont {Strand}\ \emph {et~al.}(2013)\citenamefont {Strand},
  \citenamefont {Ware}, \citenamefont {Beaudoin}, \citenamefont {Ohki},
  \citenamefont {Johnson}, \citenamefont {Blais},\ and\ \citenamefont
  {Plourde}}]{Strand.2013}%
  \BibitemOpen
  \bibfield  {author} {\bibinfo {author} {\bibfnamefont {J.~D.}\ \bibnamefont
  {Strand}}, \bibinfo {author} {\bibfnamefont {M.}~\bibnamefont {Ware}},
  \bibinfo {author} {\bibfnamefont {F.}~\bibnamefont {Beaudoin}}, \bibinfo
  {author} {\bibfnamefont {T.~A.}\ \bibnamefont {Ohki}}, \bibinfo {author}
  {\bibfnamefont {B.~R.}\ \bibnamefont {Johnson}}, \bibinfo {author}
  {\bibfnamefont {A.}~\bibnamefont {Blais}},\ and\ \bibinfo {author}
  {\bibfnamefont {B.~L.~T.}\ \bibnamefont {Plourde}},\ }\bibfield  {title}
  {\bibinfo {title} {{First-order sideband transitions with flux-driven
  asymmetric transmon qubits}},\ }\href
  {https://doi.org/10.1103/physrevb.87.220505} {\bibfield  {journal} {\bibinfo
  {journal} {Physical Review B}\ }\textbf {\bibinfo {volume} {87}},\ \bibinfo
  {pages} {220505} (\bibinfo {year} {2013})}\BibitemShut {NoStop}%
\bibitem [{\citenamefont {Blais}\ \emph {et~al.}(2007)\citenamefont {Blais},
  \citenamefont {Gambetta}, \citenamefont {Wallraff}, \citenamefont {Schuster},
  \citenamefont {Girvin}, \citenamefont {Devoret},\ and\ \citenamefont
  {Schoelkopf}}]{Blais.2007}%
  \BibitemOpen
  \bibfield  {author} {\bibinfo {author} {\bibfnamefont {A.}~\bibnamefont
  {Blais}}, \bibinfo {author} {\bibfnamefont {J.}~\bibnamefont {Gambetta}},
  \bibinfo {author} {\bibfnamefont {A.}~\bibnamefont {Wallraff}}, \bibinfo
  {author} {\bibfnamefont {D.~I.}\ \bibnamefont {Schuster}}, \bibinfo {author}
  {\bibfnamefont {S.~M.}\ \bibnamefont {Girvin}}, \bibinfo {author}
  {\bibfnamefont {M.~H.}\ \bibnamefont {Devoret}},\ and\ \bibinfo {author}
  {\bibfnamefont {R.~J.}\ \bibnamefont {Schoelkopf}},\ }\bibfield  {title}
  {\bibinfo {title} {{Quantum-information processing with circuit quantum
  electrodynamics}},\ }\href {https://doi.org/10.1103/physreva.75.032329}
  {\bibfield  {journal} {\bibinfo  {journal} {Physical Review A}\ }\textbf
  {\bibinfo {volume} {75}},\ \bibinfo {pages} {032329} (\bibinfo {year}
  {2007})}\BibitemShut {NoStop}%
\bibitem [{\citenamefont {Galliou}\ \emph {et~al.}(2013)\citenamefont
  {Galliou}, \citenamefont {Goryachev}, \citenamefont {Bourquin}, \citenamefont
  {Abbé}, \citenamefont {Aubry},\ and\ \citenamefont {Tobar}}]{Galliou.2013}%
  \BibitemOpen
  \bibfield  {author} {\bibinfo {author} {\bibfnamefont {S.}~\bibnamefont
  {Galliou}}, \bibinfo {author} {\bibfnamefont {M.}~\bibnamefont {Goryachev}},
  \bibinfo {author} {\bibfnamefont {R.}~\bibnamefont {Bourquin}}, \bibinfo
  {author} {\bibfnamefont {P.}~\bibnamefont {Abbé}}, \bibinfo {author}
  {\bibfnamefont {J.~P.}\ \bibnamefont {Aubry}},\ and\ \bibinfo {author}
  {\bibfnamefont {M.~E.}\ \bibnamefont {Tobar}},\ }\bibfield  {title} {\bibinfo
  {title} {{Extremely Low Loss Phonon-Trapping Cryogenic Acoustic Cavities for
  Future Physical Experiments}},\ }\href {https://doi.org/10.1038/srep02132}
  {\bibfield  {journal} {\bibinfo  {journal} {Scientific Reports}\ }\textbf
  {\bibinfo {volume} {3}},\ \bibinfo {pages} {2132} (\bibinfo {year}
  {2013})}\BibitemShut {NoStop}%
\bibitem [{\citenamefont {Kharel}\ \emph {et~al.}(2018)\citenamefont {Kharel},
  \citenamefont {Chu}, \citenamefont {Power}, \citenamefont {Renninger},
  \citenamefont {Schoelkopf},\ and\ \citenamefont {Rakich}}]{Kharel.20188in}%
  \BibitemOpen
  \bibfield  {author} {\bibinfo {author} {\bibfnamefont {P.}~\bibnamefont
  {Kharel}}, \bibinfo {author} {\bibfnamefont {Y.}~\bibnamefont {Chu}},
  \bibinfo {author} {\bibfnamefont {M.}~\bibnamefont {Power}}, \bibinfo
  {author} {\bibfnamefont {W.~H.}\ \bibnamefont {Renninger}}, \bibinfo {author}
  {\bibfnamefont {R.~J.}\ \bibnamefont {Schoelkopf}},\ and\ \bibinfo {author}
  {\bibfnamefont {P.~T.}\ \bibnamefont {Rakich}},\ }\bibfield  {title}
  {\bibinfo {title} {{Ultra-high-Q phononic resonators on-chip at cryogenic
  temperatures}},\ }\href {https://doi.org/10.1063/1.5026798} {\bibfield
  {journal} {\bibinfo  {journal} {APL Photonics}\ }\textbf {\bibinfo {volume}
  {3}},\ \bibinfo {pages} {066101} (\bibinfo {year} {2018})}\BibitemShut
  {NoStop}%
\bibitem [{\citenamefont {González Díaz-Palacio}\ \emph
  {et~al.}(2023)\citenamefont {González Díaz-Palacio}, \citenamefont
  {Wenskat}, \citenamefont {Deyu}, \citenamefont {Hillert}, \citenamefont
  {Blick},\ and\ \citenamefont {Zierold}}]{Gonzalez.2023}%
  \BibitemOpen
  \bibfield  {author} {\bibinfo {author} {\bibfnamefont {I.}~\bibnamefont
  {González Díaz-Palacio}}, \bibinfo {author} {\bibfnamefont
  {M.}~\bibnamefont {Wenskat}}, \bibinfo {author} {\bibfnamefont {G.~K.}\
  \bibnamefont {Deyu}}, \bibinfo {author} {\bibfnamefont {W.}~\bibnamefont
  {Hillert}}, \bibinfo {author} {\bibfnamefont {R.~H.}\ \bibnamefont {Blick}},\
  and\ \bibinfo {author} {\bibfnamefont {R.}~\bibnamefont {Zierold}},\
  }\bibfield  {title} {\bibinfo {title} {{Thermal annealing of superconducting
  niobium titanium nitride thin films deposited by plasma-enhanced atomic layer
  deposition}},\ }\href {https://doi.org/10.1063/5.0155557} {\bibfield
  {journal} {\bibinfo  {journal} {Journal of Applied Physics}\ }\textbf
  {\bibinfo {volume} {134}},\ \bibinfo {pages} {035301} (\bibinfo {year}
  {2023})}\BibitemShut {NoStop}%
\bibitem [{\citenamefont {Valery}\ \emph {et~al.}(2022)\citenamefont {Valery},
  \citenamefont {Chowdhury}, \citenamefont {Jones},\ and\ \citenamefont
  {Didier}}]{Valery.2022}%
  \BibitemOpen
  \bibfield  {author} {\bibinfo {author} {\bibfnamefont {J.~A.}\ \bibnamefont
  {Valery}}, \bibinfo {author} {\bibfnamefont {S.}~\bibnamefont {Chowdhury}},
  \bibinfo {author} {\bibfnamefont {G.}~\bibnamefont {Jones}},\ and\ \bibinfo
  {author} {\bibfnamefont {N.}~\bibnamefont {Didier}},\ }\bibfield  {title}
  {\bibinfo {title} {Dynamical sweet spot engineering via two-tone flux
  modulation of superconducting qubits},\ }\href
  {https://doi.org/10.1103/PRXQuantum.3.020337} {\bibfield  {journal} {\bibinfo
   {journal} {PRX Quantum}\ }\textbf {\bibinfo {volume} {3}},\ \bibinfo {pages}
  {020337} (\bibinfo {year} {2022})}\BibitemShut {NoStop}%
\bibitem [{\citenamefont {Law}\ and\ \citenamefont {Eberly}(1996)}]{Law.1996}%
  \BibitemOpen
  \bibfield  {author} {\bibinfo {author} {\bibfnamefont {C.~K.}\ \bibnamefont
  {Law}}\ and\ \bibinfo {author} {\bibfnamefont {J.~H.}\ \bibnamefont
  {Eberly}},\ }\bibfield  {title} {\bibinfo {title} {Arbitrary control of a
  quantum electromagnetic field},\ }\href
  {https://doi.org/10.1103/PhysRevLett.76.1055} {\bibfield  {journal} {\bibinfo
   {journal} {Physical Review Letters}\ }\textbf {\bibinfo {volume} {76}},\
  \bibinfo {pages} {1055} (\bibinfo {year} {1996})}\BibitemShut {NoStop}%
\bibitem [{\citenamefont {Krämer}\ \emph {et~al.}(2018)\citenamefont
  {Krämer}, \citenamefont {Plankensteiner}, \citenamefont {Ostermann},\ and\
  \citenamefont {Ritsch}}]{Kramer.2018}%
  \BibitemOpen
  \bibfield  {author} {\bibinfo {author} {\bibfnamefont {S.}~\bibnamefont
  {Krämer}}, \bibinfo {author} {\bibfnamefont {D.}~\bibnamefont
  {Plankensteiner}}, \bibinfo {author} {\bibfnamefont {L.}~\bibnamefont
  {Ostermann}},\ and\ \bibinfo {author} {\bibfnamefont {H.}~\bibnamefont
  {Ritsch}},\ }\bibfield  {title} {\bibinfo {title} {{QuantumOptics.jl: A Julia
  framework for simulating open quantum systems}},\ }\href
  {https://doi.org/10.1016/j.cpc.2018.02.004} {\bibfield  {journal} {\bibinfo
  {journal} {Computer Physics Communications}\ }\textbf {\bibinfo {volume}
  {227}},\ \bibinfo {pages} {109} (\bibinfo {year} {2018})}\BibitemShut
  {NoStop}%
\end{thebibliography}%


\begin{thebibliography}{5}
\expandafter\ifx\csname natexlab\endcsname\relax\def\natexlab#1{#1}\fi
\expandafter\ifx\csname bibnamefont\endcsname\relax
  \def\bibnamefont#1{#1}\fi
\expandafter\ifx\csname bibfnamefont\endcsname\relax
  \def\bibfnamefont#1{#1}\fi
\expandafter\ifx\csname citenamefont\endcsname\relax
  \def\citenamefont#1{#1}\fi
\expandafter\ifx\csname url\endcsname\relax
  \def\url#1{\texttt{#1}}\fi
\expandafter\ifx\csname urlprefix\endcsname\relax\def\urlprefix{URL }\fi
\providecommand{\bibinfo}[2]{#2}
\providecommand{\eprint}[2][]{\url{#2}}

\bibitem[{\citenamefont{Girvin}(2011)}]{Girvin.2011}
\bibinfo{author}{\bibfnamefont{S.~M.} \bibnamefont{Girvin}},
  \bibinfo{journal}{École d’été Les Houches}  (\bibinfo{year}{2011}).

\bibitem[{\citenamefont{Vool and Devoret}(2017)}]{Vool.2017}
\bibinfo{author}{\bibfnamefont{U.}~\bibnamefont{Vool}} \bibnamefont{and}
  \bibinfo{author}{\bibfnamefont{M.}~\bibnamefont{Devoret}},
  \bibinfo{journal}{International Journal of Circuit Theory and Applications}
  \textbf{\bibinfo{volume}{45}}, \bibinfo{pages}{897} (\bibinfo{year}{2017}).

\bibitem[{\citenamefont{Ambegaokar and Baratoff}(1963)}]{Ambegaokar.1963}
\bibinfo{author}{\bibfnamefont{V.}~\bibnamefont{Ambegaokar}} \bibnamefont{and}
  \bibinfo{author}{\bibfnamefont{A.}~\bibnamefont{Baratoff}},
  \bibinfo{journal}{Physical Review Letters} \textbf{\bibinfo{volume}{10}},
  \bibinfo{pages}{486} (\bibinfo{year}{1963}).

\bibitem[{\citenamefont{Martinis}(2004)}]{Martinis.2004}
\bibinfo{author}{\bibfnamefont{J.~M.} \bibnamefont{Martinis}},
  \bibinfo{journal}{Les Houches} \textbf{\bibinfo{volume}{79}},
  \bibinfo{pages}{487} (\bibinfo{year}{2004}).

\bibitem[{\citenamefont{Landauer}(1957)}]{Landauer.1957}
\bibinfo{author}{\bibfnamefont{R.}~\bibnamefont{Landauer}},
  \bibinfo{journal}{IBM Journal of Research and Development}
  \textbf{\bibinfo{volume}{1}}, \bibinfo{pages}{223} (\bibinfo{year}{1957}).

\end{thebibliography}

\end{document}

%\preprint{APS/123-QED}

\title{Hybrid optomechanical superconducting qubit system -- 
SUPPLEMENTAL MATERIAL}% Force line breaks with \\
%\thanks{A footnote to the article title}%

\author{Juuso Manninen}
\affiliation{Department of Science and Industry Systems, University of South-Eastern Norway, PO Box 235, Kongsberg, Norway
}%
 %\altaffiliation[Also at ]{Physics Department, XYZ University.}%Lines break automatically or can be forced with \\
\author{Robert H. Blick}%
\affiliation{%
 Center for Hybrid Nanostructures (CHyN), Universit\"at Hamburg,
Luruper Chaussee 149, Hamburg 22761 Germany
}%
\affiliation{
 Materials Science and Engineering, University of Wisconsin-Madison,
 1509 University Ave. WI 53706 U.S.A.
}%
\author{Francesco Massel}
\email{francesco.massel@usn.no}
\affiliation{%
 Department of Science and Industry Systems, University of South-Eastern Norway, PO Box 235, Kongsberg, Norway
}%

%\date{\today}% It is always \today, today,
             %  but any date may be explicitly specified

%\begin{abstract}
%\begin{description}
%\item[Usage]
%Secondary publications and information retrieval purposes.
%\item[Structure]
%You may use the \texttt{description} environment to structure your abstract;
%use the optional argument of the \verb+\item+ command to give the category of each item. 
%\end{description}
%\end{abstract}

%\keywords{Suggested keywords}%Use showkeys class option if keyword
                              %display desired
\maketitle

%\tableofcontents

%xxxxxxxxxxxxxxxxxxxxxxxxxxxxxxxxxxxx
% DERIVATION OF THE HAMILTONIAN
%xxxxxxxxxxxxxxxxxxxxxxxxxxxxxxxxxxxx
\section{Derivation of the X2MON Hamiltonian}

We present here a full derivation of the Hamiltonian of the X2MON 
circuit (see Fig. \ref{fig:SI1}) and of the couplings between the mechanical motion 
and the qubit that arise in the system. The calculation here generalizes the 
discussion of the main text, we do not assume the Josephson junctions 
in the system to be symmetric. Note that we use the convention $\hbar = 1$.

\subsection{Lumped-element description}
\begin{figure}[htb]
  \includegraphics[width=0.9\textwidth]{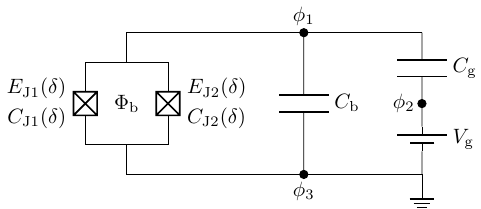}
  \caption{Schematic of the SUNMESH circuit. For the formal calculation, 
  the voltage source is replaced by a capacitor $\cvg$ allowing voltage 
  $V_\mathrm{g}$ to be induced on island 2.}
  \label{fig:SI1}
\end{figure}

We derive the Hamiltonian for the X2MON circuit with standard circuit quantum 
electrodynamics (circuit-QED) methods~\cite{Girvin.2011,Vool.2017}. 
The flux at the node $i$ at time $t$ is given by
\begin{equation}
\label{eq:SI1}
    \Phi_i (t) = \int^t V_i(\tau) d\tau ,
\end{equation}
which defines the voltage of the node as 
\begin{equation}
\label{eq:SI2}
    V_i(t) = \dot{\Phi}_i .
\end{equation}
The node flux and phase are connected with the following relation
\begin{equation}
\label{eq:SI3}
    \phi_i = 2\pi \frac{\Phi_i}{\Phi_0} ,
\end{equation}
where $\Phi_0 = \frac{h}{2e}$ is the flux quantum. The indices of different nodes 
in our circuit are given in Fig. \ref{fig:SI1}.

The energy related to the capacitive elements of the circuit is given by
\begin{equation}
\label{eq:SI4}
    \mathcal{T} = \frac{\cjone + \cjtwo + \cb}{2} \dot{\Phi}_1^2 
        + \frac{\cg}{2} ( \dot{\Phi}_1 - \dot{\Phi}_2 )^2 
        + \frac{\cvg}{2} \dot{\Phi}_2^2 .
\end{equation}
Here $\cjone$ and $\cjtwo$ are the capacitances of the Josephson junctions, $\cg$ is 
the gate capacitance, and $\cb$ the shunting capacitance that is large so that 
our qubit operates in the TRANSMON regime. $\cvg$ is an additional capacitance,
replacing the voltage source for the formal calculation, 
whose only function in this formalism is to induce the voltage bias on the qubit island, 
see Ref.~\cite{Girvin.2011}. Note that $\dot{\Phi}_3$ is not present here, since 
$\dot{\Phi}_3 = 0$ is due to that node being grounded.
Only the Josephson junctions contribute to the inductive energy of the circuit,
\begin{equation}
\label{eq:SI5}
\begin{split}
    \mathcal{U} &= -\ejone (\delta) \cos \left( 2\pi \frac{\Phi_1 + \frac{1}{2}\Phi_\mathrm{b}}{\Phi_0} \right) 
            -\ejtwo (\delta) \cos \left( 2\pi \frac{\Phi_1 - \frac{1}{2}\Phi_\mathrm{b}}{\Phi_0} \right) \\
        &= -\left(\ejone (\delta) + \ejtwo (\delta) \right) \left( \cos \phi_1 \cos \frac{\phib}{2} 
            + d(\delta) \sin \phi_1 \sin \frac{\phib}{2} \right) ,
\end{split}
\end{equation}
where $d(\delta) = (\ejtwo (\delta) - \ejone (\delta))/(\ejone (\delta) + \ejtwo(\delta))$ 
and the Josephson energies are given by
\begin{subequations}
\label{eq:SI6}
\begin{align}
    \ejone (\delta) &= \tilde{E}_{\mathrm{J}10} e^{-\frac{x_0 + \delta}{\xi}} 
            = \ejonez e^{-\frac{\delta}{\xi}} ,\\
    \ejtwo (\delta) &= \tilde{E}_{\mathrm{J}20} e^{-\frac{x_0 - \delta}{\xi}}
            = \ejtwoz e^{\frac{\delta}{\xi}} .
\end{align}
\end{subequations}
Anticipating our analysis of the device dynamics, in
Eqs.~(\ref{eq:SI5},\ref{eq:SI6}), we have explicitly indicated the JJ
energy dependence on the deviation ($\delta$) from the equilibrium
position($x_0$) of the shuttle.
\subsection{Josephson energy}

We provide here a brief explanation of the forms of the Josephson
energies of the two junctions given in Eq.~\eqref{eq:SI6}.  Let us
assume that the normal state resistance of the junctions is
\begin{equation}
\label{eq:SI_RN}
    R_\mathrm{N} = R_{\mathrm{N}0} \exp \left[\frac{x}{\xi}\right] ,
\end{equation}
stemming from the exponentially suppressed tunneling probability
across the junction as a function of its thickness $x$. 
The Ambegaokar--Baratoff formula~\cite{Ambegaokar.1963,Martinis.2004}
provides the following expression for the JJ energy
\begin{equation}
\label{eq:SI_AB_formula}
    E_\mathrm{J}(\phi) = - \frac{1}{8} \frac{R_\mathrm{K}}{R_\mathrm{N}} \Delta \cos \phi ,
\end{equation}
where $R_\mathrm{K} = \frac{2\pi \hbar}{e^2}$ is the resistance quantum, 
$\Delta$ the superconducting gap, and $\phi$ the superconducting phase difference 
across the junction. Focusing on the characteristic Josephson energy 
$E_\mathrm{J} = \left| E_\mathrm{J} (\phi = 0) \right|$, we get
Eq.~\eqref{eq:SI6} 
with the help of the Landauer formula~\cite{Landauer.1957} and Eq.~\eqref{eq:SI_RN},  
where the parameter $\xi$ is given by
\begin{equation}
\label{eq:SI_xi}
    \xi = \frac{x_0}{\log\left[ \frac{\Delta}{8 E_{\mathrm{J}}} 
                    \frac{R_\mathrm{K}}{R_{\mathrm{N}0}}\right]}.
\end{equation}
\subsection{X2MON Hamiltonian}
The Lagrangian of the system, including the  Lagrangian of the
shuttle $\mathcal{L}_\mathrm{m} \left(x_\mathrm{m},p_\mathrm{m}\right)$,
\begin{equation}
\label{eq:SI7}
    \mathcal{L} = \mathcal{T} - \mathcal{U} = \frac{1}{2} \vec{\dot{\Phi}}^\top \left[ C \right] \vec{\dot{\Phi}} - \mathcal{U} + \mathcal{L}_\mathrm{m} \left(x_\mathrm{m},p_\mathrm{m}\right)
\end{equation}
can be written using the capacitance matrix
\begin{equation}
\label{eq:SI8}
    \left[ C \right] = 
    \begin{pmatrix}
        \cjone + \cjtwo + \cb + \cg & -\cg \\
        -\cg & \cg + \cvg
    \end{pmatrix}
\end{equation}
whose inverse is 
\begin{equation}
\label{eq:SI9}
    \left[ C \right]^{-1} = 
    \frac{1}{(\cg + \cb)(\cg + \cvg) + \cg \cvg}
    \begin{pmatrix}
        \cg + \cvg & \cg \\
        \cg & \cg + \cb + \cg
    \end{pmatrix}
    = 
    \begin{pmatrix}
        \frac{1}{C_{11}} & \frac{1}{C_{12}} \\
        \frac{1}{C_{12}} & \frac{1}{C_{22}}
    \end{pmatrix} .
\end{equation}
The Hamiltonian can be expressed using this inverse capacitance matrix
\begin{equation}
\label{eq:SI10}
\begin{split}
    \mathcal{H} &= \sum_i \dot{\Phi}_i \frac{\partial \mathcal{L}}{\partial \dot{\Phi}_i} + \dot{x}_\mathrm{m} \frac{\partial \mathcal{L}}{\partial \dot{x}_\mathrm{m}} - \mathcal{L} \\
        &= \sum_i \dot{\Phi}_i Q_i + \dot{x}_\mathrm{m} \frac{\partial \mathcal{L}}{\partial \dot{x}_\mathrm{m}} - \mathcal{L} \\
        &= \frac{1}{2} \vec{Q}^\top \left[ C \right]^{-1} \vec{Q} + \mathcal{U} + E\left(x_\mathrm{m},p_\mathrm{m}\right) \\
        &= \frac{1}{2 C_{11}} Q_1^2 + \frac{1}{C_{12}} Q_1 Q_2 + \frac{1}{2 C_{22}} Q_2^2 + \mathcal{U} + E\left(x_\mathrm{m},p_\mathrm{m}\right),
\end{split}
\end{equation}
where $E\left(x_\mathrm{m},p_\mathrm{m}\right)$ is the (elastic)
energy of the shuttle, and the conjugate variable to flux $\Phi_i$ are charges on each island 
$Q_i = \frac{\partial \mathcal{L}}{\partial \dot{\Phi}_i}$. 
Note that the voltage on island 2, $\frac{\partial \mathcal{H}}{\partial Q_2}$, 
can now be set to the gate voltage $V_\mathrm{g}$ by choosing 
$V_\mathrm{g} = \frac{Q_2}{C_{22}}$ and letting $\cvg \to \infty$.

Now, defining the number of Cooper pairs on the qubit island, the charging energy of the qubit, 
and the gate charge
\begin{equation}
\label{eq:SI11}
\begin{split}
    & n = \frac{Q_1}{2e} ,\\
    & \ec (\delta) = \frac{e^2}{2 C_{11}} 
            \xrightarrow{ \cvg \to \infty } 
                \frac{e^2}{2 ( \cjone (\delta) + \cjtwo (\delta) + \cb + \cg )} ,\\
    & n_\mathrm{g} = - \frac{1}{2e} \frac{C_{11} C_{22}}{C_{12}} V_\mathrm{g} 
            \xrightarrow{ \cvg \to \infty } - \frac{\cg}{2e} V_\mathrm{g} ,
\end{split}
\end{equation}
the Hamiltonian in Eq. \eqref{eq:SI10} can be written in the canonical form
\begin{equation}
\label{eq:SI12}
    \mathcal{H} = 4 \ec (\delta) ( n - n_\mathrm{g} )^2  + \mathcal{U} 
             + E\left(x_\mathrm{m},p_\mathrm{m}\right)
             + 2 e^2 n_\mathrm{g}^2 \left( \frac{C_{12}^2}{C_{11}^2 C_{22}} - \frac{1}{C_{11}} \right) ,
\end{equation}
where the last term can be ignored due to it providing only a constant contribution to the 
total energy of the system and not affecting its dynamics.

%xxxxxxxxxxxxxxxxxxxxxxxxxxxxxxxxxxxx
% QUANTIZATION OF THE HAMILTONIAN
%xxxxxxxxxxxxxxxxxxxxxxxxxxxxxxxxxxxx
\subsection{Quantization of the Hamiltonian}

Firstly, we expand the potential energy part of the Hamiltonian 
in Eq.~\eqref{eq:SI12} up to the fourth order in $\phi_1$
\begin{equation}
\label{eq:SI13}
\begin{split}
    \mathcal{H} \approx& \ 4 \ec (\delta) ( n - n_\mathrm{g} )^2 + E\left(x_\mathrm{m},p_\mathrm{m}\right) \\
        &- \left(\ejone (\delta) + \ejtwo (\delta) \right) \cos \frac{\phib}{2} 
            \left( 1 - \frac{\phi_1^2}{2} + \frac{\phi_1^4}{24}\right) \\
        &- \left(\ejone (\delta) + \ejtwo (\delta) \right) d(\delta) \sin \frac{\phib}{2} 
            \left( \phi_1 - \frac{\phi_1^3}{6}\right) .
\end{split} 
\end{equation}
We then promote the phase $\phi$ and $n$ to quantum operators
$\hat{\phi}$ and $\hat{n}$, respectively. These obey canonical
commutation relations
\begin{equation}
\label{eq:SI14}
    \left[ \hat{\phi}_1 , \hat{n} \right] = i ,
\end{equation}
%
%arising from the canonical commutation relation of conjugate variables 
%$\left[ \hat{\Phi}_1 , \hat{Q}_1 \right] = i$.
We can express them in terms of bosonic lowering (raising) operators
$a$ $(a^\dagger)$, which obey the standard bosonic commutation relation
$\left[ a , a^\dagger \right] = 1$, as
\begin{subequations}
\label{eq:SI15}
\begin{align}
    \hat{\phi}_1 &= \phizpf ( a^\dagger + a ) ,\\
    \hat{n} &= i \nzpf ( a^\dagger - a ). 
\end{align}
\end{subequations}
Substituting these definitions into Eq.~\eqref{eq:SI14} implies for 
the zero-point fluctuations
\begin{equation}
\label{eq:SI16}
    \phizpf \nzpf = \frac{1}{2} .
\end{equation}
The quantized form of the Hamiltonian, Eq.~\eqref{eq:SI13}, can then be written as
\begin{equation}
\label{eq:SI17}
\begin{split}
    \hat{H} =& 
        \left( 4 \ec (\delta) \nzpf^2 
            + \frac{1}{2} \left(\ejone (\delta) + \ejtwo (\delta) \right) \cos \frac{\phib}{2} \phizpf^2 \right)
            ( 2 a^\dagger a + 1) \\
        &+ \left( -4 \ec (\delta) \nzpf^2 
            + \frac{1}{2} \left(\ejone (\delta) + \ejtwo (\delta) \right) \cos \frac{\phib}{2} \phizpf^2 \right)
            ( a^{\dagger 2} + a^2 ) \\
        &- \left(\ejone (\delta) + \ejtwo (\delta) \right) \cos \frac{\phib}{2} 
            \left( 1 + \frac{\hat{\phi}^4}{24}\right) \\
        &- \left(\ejone (\delta) + \ejtwo (\delta) \right) d(\delta) \sin \frac{\phib}{2} 
            \left( \hat{\phi} - \frac{\hat{\phi}^3}{6}\right) \\
        &+ E\left(x_0+\delta,p_\mathrm{m}\right).
\end{split}
\end{equation}
Within the conventional harmonic description \cite{Girvin.2011},  
recalling the identity Eq.~\eqref{eq:SI16}, we can express the
zero-point fluctuations as
\begin{equation}
\label{eq:SI18}
\begin{split}
    \nzpf (\delta) &= \left\{ \frac{\ejone (\delta) + \ejtwo (\delta)}{32 \ec (\delta)}  \cos \frac{\phib}{2} \right\}^{\frac{1}{4}}, \\
    \phizpf (\delta) &= \left\{ \frac{2 \ec (\delta)}
        {\left( \ejone (\delta) + \ejtwo (\delta) \right) \cos \frac{\phib}{2} }  \right\}^{\frac{1}{4}},
\end{split}
\end{equation}
and the full expression of the Hamiltonian becomes
\begin{equation}
\label{eq:SI19}
\begin{split}
    \hat{H} =& \ \wpp (\delta) ( a^\dagger a + \frac{1}{2} ) + E\left(x_0+\delta,p_\mathrm{m}\right)\\
        &- \left(\ejone (\delta) + \ejtwo (\delta) \right) \cos \frac{\phib}{2} 
        - \frac{1}{12} \ec (\delta) ( a^\dagger + a )^4 \\
        &- \left(\ejone (\delta) + \ejtwo (\delta) \right) d (\delta) 
            \sin \frac{\phib}{2} \phizpf (\delta) ( a^\dagger + a) \\
        &+ \frac{1}{6} \left(\ejone (\delta) + \ejtwo (\delta) \right) d (\delta) 
            \sin \frac{\phib}{2} \phizpf^3 (\delta) ( a^\dagger + a)^3 ,
\end{split}
\end{equation}
where
\begin{equation}
\label{eq:SI20}
    \wpp (\delta) = \sqrt{8 \ec (\delta) \left(\ejone (\delta) + \ejtwo (\delta) \right) 
        \cos \frac{\phib}{2}} . 
\end{equation}
%

%xxxxxxxxxxxxxxxxxxxxxxxxxxxxxxxxxxxx
% INTERACTION BETWEEN THE QUBIT AND THE MECHANICS
%xxxxxxxxxxxxxxxxxxxxxxxxxxxxxxxxxxxx
\section{Deriving the coupling between the qubit and the mechanics}

We now expand the Hamiltonian in Eq.~\eqref{eq:SI19} with respect to
the mechanical displacement up to the second order and denote the
quantized mechanical displacement (obeying canonical commutation
relations) with $\hat{\delta} = \xzpf (b^\dagger + b)$.  The following
shorthand notation is used to display the results in a more compact
form: $\ecz = \ec (0)$, $\ejonez = \ejone (0)$,
$\ejtwoz = \ejtwo (0)$, and $d_0 = d(0)$.  The Josephson capacitances
are approximated with parallel-plate capacitors, i.e. the charging
energy has the form
\begin{equation}
\label{eq:SI21}
    \ec (\delta) = \frac{e^2}{2 \left( 
    \frac{C_{\mathrm{J}}}{1+\frac{\delta}{x_0}} + \frac{C_{\mathrm{J}}}{1-\frac{\delta}{x_0}}
    + \cb + \cg \right)} ,
\end{equation}
where $C_{\mathrm{J}} = \cjone (0) = \cjtwo (0)$.

Let us go through the expansion of the Hamiltonian,
Eq.~\eqref{eq:SI19}, term by term. The coefficient definitions are
summarized below in Eq.~\eqref{eq:SI30} for the reader's convenience. 

For the $(a^\dagger a + \frac{1}{2})$ term, we obtain
\begin{equation}
\label{eq:SI22}
\begin{split}
    &\wpp (\hat{\delta}) ( a^\dagger a + \frac{1}{2} ) \\
    &\approx 
    \left\{
    \sqrt{8 \ecz (\ejonez + \ejtwoz) \cos \frac{\phib}{2}} + 
        d_0 \sqrt{2 \ecz (\ejonez + \ejtwoz) \cos \frac{\phib}{2}} \frac{\xzpf}{\xi} (b^\dagger + b) \right. \\
        & \left. \quad + \sqrt{8 \ecz (\ejonez + \ejtwoz) \cos \frac{\phib}{2}} 
            \left[ \frac{1}{2 \xi^2} - \frac{2 \cjzero}{(2 \cjzero + \cb + \cg) x_0^2} 
                - \frac{d_0^2}{4 \xi^2} \right] \xzpf^2 (b^\dagger + b)^2 
                \right\} \\
        &\qquad \times ( a^\dagger a + \frac{1}{2} ) \\
        &\approx 
        \left\{
        \sqrt{8 \ecz (\ejonez + \ejtwoz) \cos \frac{\phib}{2}} + 
        d_0 \sqrt{2 \ecz (\ejonez + \ejtwoz) \cos \frac{\phib}{2}} \frac{\xzpf}{\xi} (b^\dagger + b) \right. \\
        &\left. \quad + \frac{1}{2} \sqrt{2 \ecz (\ejonez + \ejtwoz) \cos \frac{\phib}{2}} 
             \left(\frac{\xzpf}{\xi}\right)^2 (b^\dagger + b)^2 
        \right\} ( a^\dagger a + \frac{1}{2} ) \\
        &= \left\{ 
        \wpzero + g_{21} (b^\dagger + b) + g_{22} (b^\dagger + b)^2 
        \right\} ( a^\dagger a + \frac{1}{2} ),
\end{split}
\end{equation}
and the term related to $(a^\dagger + a)^4$ gives
\begin{equation}
\label{eq:SI23}
\begin{split}
    - \frac{\ec (\hat{\delta})}{12} (a^\dagger + a)^4 
    &\approx 
    \left\{
    - \frac{\ecz}{12} 
    + \frac{1}{3 e^2} \ecz^2 \cjzero \left(\frac{\xzpf}{\xi}\right)^2 (b^\dagger + b)^2 
    \right\} (a^\dagger + a)^4 \\
    &= \left\{
    - \frac{\ecz}{12} + g_{42} (b^\dagger + b)^2 
    \right\} (a^\dagger + a)^4,
\end{split}
\end{equation}
where the first term contributes to the renormalization of the qubit frequency.

The  energy of the shuttle $E\left(x_0+\delta,p_\mathrm{m}\right)$ is
expressed as the sum of a kinetic $p_\mathrm{m}/2m$ and an elastic
$1/2 \,m \omega_\mathrm{m}^2\left(x_0+\delta\right)^2 $ contribution; it is quantized as
$\omega_\mathrm{m0} b^\dagger b$ 
where $\omega_\mathrm{m0}$ is the bare mechanical frequency of the
shuttle. The frequency $\omega_\mathrm{m0}$ is renormalized partly by 
the contribution proportional to $b^\dagger b$ from the following term 
\begin{equation}
\label{eq:SI24}
\begin{split}
    - (\ejone (\hat{\delta}) + \ejtwo (\hat{\delta}) ) \cos \frac{\phib}{2} 
                &\approx - (\ejonez + \ejtwoz) \cos \frac{\phib}{2} \\
                    &\quad - (\ejtwoz - \ejonez) \cos \frac{\phib}{2} \frac{\xzpf}{\xi} (b^\dagger + b) \\
                    &\quad - \frac{1}{2} (\ejonez + \ejtwoz) \cos \frac{\phib}{2}
                        \left(\frac{\xzpf}{\xi}\right)^2 (b^\dagger + b)^2 \\
                &= - (\ejonez + \ejtwoz) \cos \frac{\phib}{2} + g_{01} (b^\dagger + b) 
                    + g_{02} (b^\dagger + b)^2,
\end{split}
\end{equation}
arising from the coupling to the qubit.
% Above, the even exponents in $a$ and $a^\dagger$ contribute to the renormalization 
% of the qubit and mechanical frequencies, and below the terms with odd exponents are 
% related to transverse qubit terms.
We now expand the coefficients appearing in Eq.~\eqref{eq:SI19}.
The coefficient corresponding to the $(a^\dagger + a)$ term is
  \begin{equation*}
    \begin{split}
      &\left\{
      - (\ejone (\hat{\delta}) + \ejtwo (\hat{\delta}) ) d (\hat{\delta}) 
        \sin \frac{\phib}{2} \phizpf (\hat{\delta}) 
        \right\} (a^\dagger + a) \\
      &\approx 
      \left\{
      - ( \ejtwoz - \ejonez ) 
        \left( \frac{2 \ecz}{(\ejonez + \ejtwoz) \cos \frac{\phib}{2}} \right)^{\frac{1}{4}} 
        \sin \frac{\phib}{2} 
        \right. \\
      &\quad - \frac{(3 \ejonez + \ejtwoz)(\ejonez + 3 \ejtwoz)}{4 (\ejonez + \ejtwoz)} 
        \left( \frac{2 \ecz}{(\ejonez + \ejtwoz) \cos \frac{\phib}{2}} \right)^{\frac{1}{4}} 
        \sin \frac{\phib}{2} \frac{\xzpf}{\xi} (b^\dagger + b) \\
      &\quad - \frac{(\ejtwoz - \ejonez)}{32 (\ejonez + \ejtwoz)^2} 
        ( 9 \ejonez^2 - 2 \ejonez \ejtwoz + 9 \ejtwoz^2) 
        \left( \frac{2 \ecz}{(\ejonez + \ejtwoz) \cos \frac{\phib}{2}} \right)^{\frac{1}{4}} \\
      &\qquad 
      \left. 
      \times \sin \frac{\phib}{2} \left(\frac{\xzpf}{\xi}\right)^2 (b^\dagger + b)^2
      \right\} (a^\dagger + a)
    \end{split}
  \end{equation*}         
  \begin{equation}
    \label{eq:SI25}
    \begin{split}          
      &= \left\{
      - ( \ejtwoz - \ejonez ) 
        \left( \frac{2 \ecz}{\ejonez + \ejtwoz} \right)^{\frac{1}{4}} 
        \tan \frac{\phib}{2} \cos^{\frac{3}{4}} \frac{\phib}{2} 
        \right. \\
      &\quad - \frac{(3 \ejonez + \ejtwoz)(\ejonez + 3 \ejtwoz)}{4 (\ejonez + \ejtwoz)} 
        \left( \frac{2 \ecz}{\ejonez + \ejtwoz} \right)^{\frac{1}{4}} 
        \tan \frac{\phib}{2} \cos^{\frac{3}{4}} \frac{\phib}{2} \frac{\xzpf}{\xi} (b^\dagger + b) \\
      &\quad - \frac{(\ejtwoz - \ejonez)}{32 (\ejonez + \ejtwoz)^2} 
        ( 9 \ejonez^2 - 2 \ejonez \ejtwoz + 9 \ejtwoz^2) 
        \left( \frac{2 \ecz}{\ejonez + \ejtwoz} \right)^{\frac{1}{4}} \\
      &\qquad 
      \left.
      \times \tan \frac{\phib}{2} \cos^{\frac{3}{4}} \frac{\phib}{2} 
        \left(\frac{\xzpf}{\xi}\right)^2 (b^\dagger + b)^2 
        \right\} (a^\dagger + a) \\
      &= \left\{ 
      g_{10} + g_{11} (b^\dagger + b) + g_{12} (b^\dagger + b)^2 
      \right\} (a^\dagger + a),
    \end{split}
  \end{equation}
and finally the term with $(a^\dagger + a)^3$
\begin{equation*}
\begin{split}
        &\left\{ 
        \frac{1}{6} (\ejone (\hat{\delta}) + \ejtwo (\hat{\delta}) ) d (\hat{\delta}) 
        \sin \frac{\phib}{2} \phizpf (\hat{\delta})^3 
        \right\} (a^\dagger + a)^3\\
        &\approx 
        \left\{
        \frac{1}{6} ( \ejtwoz - \ejonez ) 
            \left( \frac{2 \ecz}{(\ejonez + \ejtwoz) \cos \frac{\phib}{2}} \right)^{\frac{3}{4}} 
            \sin \frac{\phib}{2} 
        \right. \\
        &\quad + \frac{( \ejonez^2 + 14 \ejonez \ejtwoz +  \ejtwoz^2)}{24 (\ejonez + \ejtwoz)} 
            \left( \frac{2 \ecz}{(\ejonez + \ejtwoz) \cos \frac{\phib}{2}} \right)^{\frac{3}{4}} 
            \sin \frac{\phib}{2} \frac{\xzpf}{\xi} (b^\dagger + b) \\
        &\quad + \frac{(\ejtwoz - \ejonez)}{192 (\ejonez + \ejtwoz)^2} 
            ( \ejonez^2 - 82 \ejonez \ejtwoz + \ejtwoz^2) 
             \left( \frac{2 \ecz}{(\ejonez + \ejtwoz) \cos \frac{\phib}{2}} \right)^{\frac{3}{4}} \\
        &\qquad 
        \left.
        \times \sin \frac{\phib}{2} \left(\frac{\xzpf}{\xi}\right)^2 (b^\dagger + b)^2
        \right\} (a^\dagger + a)^3
\end{split}
\end{equation*}         
  \begin{equation}
    \label{eq:SI26}
    \begin{split}          
        &= 
        \left\{
        \frac{1}{6} ( \ejtwoz - \ejonez ) 
            \left( \frac{2 \ecz}{\ejonez + \ejtwoz} \right)^{\frac{3}{4}} 
            \tan \frac{\phib}{2} \cos^{\frac{1}{4}} \frac{\phib}{2}
        \right. \\
        &\quad + \frac{( \ejonez^2 + 14 \ejonez \ejtwoz +  \ejtwoz^2)}{24 (\ejonez + \ejtwoz)} 
            \left( \frac{2 \ecz}{\ejonez + \ejtwoz} \right)^{\frac{3}{4}} 
            \tan \frac{\phib}{2} \cos^{\frac{1}{4}} \frac{\phib}{2} \frac{\xzpf}{\xi} (b^\dagger + b) \\
        &\quad + \frac{(\ejtwoz - \ejonez)}{192 (\ejonez + \ejtwoz)^2} 
            ( \ejonez^2 - 82 \ejonez \ejtwoz + \ejtwoz^2) 
             \left( \frac{2 \ecz}{\ejonez + \ejtwoz} \right)^{\frac{3}{4}} \\
            &\qquad 
            \left.
            \times \tan \frac{\phib}{2} \cos^{\frac{1}{4}} \frac{\phib}{2} 
            \left(\frac{\xzpf}{\xi}\right)^2 (b^\dagger + b)^2 
            \right\} (a^\dagger + a)^3\\
        &= \left\{
        g_{30} + g_{31} (b^\dagger + b) + g_{32} (b^\dagger + b)^2 
        \right\} (a^\dagger + a)^3.
\end{split}
\end{equation}
Gathering all the terms, the full Hamiltonian is
\begin{equation}
\label{eq:SI27}
\begin{split}
    \hat{H} &= \wpzero (a^\dagger a + \frac{1}{2})  
                + g_{21} (a^\dagger a + \frac{1}{2}) (b^\dagger + b) 
                + g_{22} (a^\dagger a + \frac{1}{2}) (b^\dagger + b)^2 \\
            &\quad + \omega_{\mathrm{m}0} b^\dagger b - ( \ejonez + \ejtwoz ) \sin \frac{\phib}{2} 
                + g_{01} (b^\dagger + b) + g_{02} (b^\dagger + b)^2 \\
            &\quad - \frac{\ecz}{12} (a^\dagger + a)^4 + g_{42} (a^\dagger + a)^4 (b^\dagger + b)^2       \\
            &\quad + g_{10} (a^\dagger + a) + g_{11} (a^\dagger + a) (b^\dagger + b) 
                + g_{12} (a^\dagger + a) (b^\dagger + b)^2 \\
            &\quad + g_{30} (a^\dagger + a)^3 + g_{31} (a^\dagger + a)^3 (b^\dagger + b) 
                + g_{32} (a^\dagger + a)^3 (b^\dagger + b)^2 ,
\end{split}
\end{equation}
In the following, we neglect mixing of states outside the
computational basis and therefore consider normal ordering the qubit
operators $(a^\dagger + a)^3 \to 3\, (a^\dagger + a)$ and
$(a^\dagger + a)^4 \to 12\, a^\dagger a$. A final mapping of bosonic
operators onto Pauli matrices ($a^\dagger a \to 1/2 \,(\sigma_z+1)$ and
$a^\dagger + a \to \sigma_x$) allows us to write the final form of the Hamiltonian of
the X2MON circuit as 
\begin{equation}
\label{eq:SI28}
\begin{split}
    \hat{H} &= \frac{\omega_\mathrm{q}}{2} \sigma_z + \omega_\mathrm{m} b^\dagger b 
                + g_{01} (b^\dagger + b) \\
            &\quad + \left[ g_{21} (b^\dagger + b) 
                        + ( g_{22} + 12 g_{42}) (b^\dagger + b)^2 \right] \sigma_z  \\
            &\quad + \left[ ( g_{10} + 3 g_{30} ) + ( g_{11} + 3 g_{31} ) (b^\dagger + b) 
                        + ( g_{12} + 3 g_{32}) (b^\dagger + b)^2 \right] \sigma_x 
\end{split} 
\end{equation}
exhibiting both linear and quadratic coupling of the mechanical displacement to the 
$\sigma_x$ and $\sigma_z$ terms of the qubit. The relative strengths of the couplings 
can be tuned with the flux bias $\phib$ and also already in the fabrication stage of 
the device by choosing a suitable asymmetry of the Josephson energies of the junctions. 
Notably, the quadratic coupling to the $\sigma_x$ component is weak compared to the 
other terms and can be disregarded in general. Here, the renormalized qubit and mechanical frequencies are 
\begin{equation}
 \label{eq:SI29}
\begin{split} 
    \omega_\mathrm{q} &= 2 \left(\wpzero - \ecz \right),  \\
    \omega_\mathrm{m} &= \omega_{\mathrm{m}0} + ( 2 g_{02} + g_{22} + 6 g_{42} ), 
\end{split}
\end{equation}
respectively.

For completeness, the full list of the coupling coefficients is

\begin{equation}
  \label{eq:SI30}
\begin{split}
    \wpzero &= \sqrt{8 \ecz (\ejonez + \ejtwoz) \cos \frac{\phib}{2}} \\
    g_{21} &= \frac{d_0}{2} \frac{\xzpf}{\xi} \wpzero \\
    g_{22} &= \frac{1}{4} \left(\frac{\xzpf}{\xi}\right)^2 \wpzero \\
    g_{42} &= \frac{1}{3 e^2} \ecz^2 \cjzero \left(\frac{\xzpf}{\xi}\right)^2 \\
    g_{01} &= - (\ejtwoz - \ejonez) \cos \frac{\phib}{2} \frac{\xzpf}{\xi}  \\
    g_{02} &= - \frac{1}{2} (\ejonez + \ejtwoz) \cos \frac{\phib}{2}
                        \left(\frac{\xzpf}{\xi}\right)^2 \\
    g_{10} &= - ( \ejtwoz - \ejonez ) 
                \left( \frac{2 \ecz}{\ejonez + \ejtwoz} \right)^{\frac{1}{4}} 
                \tan \frac{\phib}{2} \cos^{\frac{3}{4}} \frac{\phib}{2} \\
    g_{11} &= - \frac{(3 \ejonez + \ejtwoz)(\ejonez + 3 \ejtwoz)}{4 (\ejonez + \ejtwoz)} 
                \left( \frac{2 \ecz}{\ejonez + \ejtwoz} \right)^{\frac{1}{4}} 
                \tan \frac{\phib}{2} \cos^{\frac{3}{4}} \frac{\phib}{2} \frac{\xzpf}{\xi} \\
    g_{12} &= - \frac{(\ejtwoz - \ejonez)}{32 (\ejonez + \ejtwoz)^2} 
                ( 9 \ejonez^2 - 2 \ejonez \ejtwoz + 9 \ejtwoz^2) 
                \left( \frac{2 \ecz}{\ejonez + \ejtwoz} \right)^{\frac{1}{4}} \\
                    &\qquad \times \tan \frac{\phib}{2} \cos^{\frac{3}{4}} \frac{\phib}{2} 
                    \left(\frac{\xzpf}{\xi}\right)^2  \\
    g_{30} &= \frac{1}{6} ( \ejtwoz - \ejonez ) 
                \left( \frac{2 \ecz}{\ejonez + \ejtwoz} \right)^{\frac{3}{4}} 
                \tan \frac{\phib}{2} \cos^{\frac{1}{4}} \frac{\phib}{2} \\
    g_{31} &= \frac{( \ejonez^2 + 14 \ejonez \ejtwoz +  \ejtwoz^2)}{24 (\ejonez + \ejtwoz)} 
            \left( \frac{2 \ecz}{\ejonez + \ejtwoz} \right)^{\frac{3}{4}} 
            \tan \frac{\phib}{2} \cos^{\frac{1}{4}} \frac{\phib}{2} \frac{\xzpf}{\xi} \\
    g_{32} &= \frac{(\ejtwoz - \ejonez)}{192 (\ejonez + \ejtwoz)^2} 
            ( \ejonez^2 - 82 \ejonez \ejtwoz + \ejtwoz^2) 
             \left( \frac{2 \ecz}{\ejonez + \ejtwoz} \right)^{\frac{3}{4}} \\
            &\qquad \times \tan \frac{\phib}{2} \cos^{\frac{1}{4}} \frac{\phib}{2} 
            \left(\frac{\xzpf}{\xi}\right)^2 .
\end{split}
\end{equation}

Crucially, for symmetric Josephson junctions $( \ejonez = \ejtwoz \equiv \ej)$, many of these 
couplings vanish, and we obtain
$g_{01} = g_{10} = g_{12} = g_{21} = g_{30} = g_{32} = 0$, resulting in the Hamiltonian 
presented in the main text Eq.~(5) with the notation $g_1 \equiv g_{11} + 3 g_{31}$ and 
$g_2 \equiv g_{22} + 12 g_{42}$ related to the couplings 
$g_1 (b^\dagger + b) \sigma_x$ 
and $g_2 (b^\dagger + b)^2 \sigma_z$.
For symmetric junctions, the coefficients to these couplings are expressed as 
\begin{equation}
\begin{split}
    g_1 &= 2 E_\mathrm{J} \left( \frac{\xzpf}{\xi} \right) \tan \frac{\phib}{2}
            \left[
                - \left( \frac{\ecz}{E_\mathrm{J}} \right)^{\frac{1}{4}} 
                \cos^{\frac{3}{4}} \frac{\phib}{2} 
                + \left( \frac{\ecz}{E_\mathrm{J}} \right)^{\frac{3}{4}} 
                \cos^{\frac{1}{4}} \frac{\phib}{2} 
                \right], \\
    g_2 &= \left(\frac{\xzpf}{\xi}\right)^2 
            \left[
            \sqrt{\ecz E_\mathrm{J} \cos \frac{\phib}{2}}
            + \frac{1}{3 e^2} \ecz^2 \cjzero
            \right] .
\end{split}
\end{equation}

\section{Approximate State-swapping Hamiltonian}

We derive here the Hamiltonian $H'$ given in Eq.~(7) of the main text,
by considering the effect of a flux modulation given by
$\phi_\mathrm{b}=\phi_\mathrm{b0} \cos\left(\bar{\omega} t\right)$.
As a result, the numerical factors appearing in the Hamiltonian given
in Eq.~(5) of the main text
\begin{equation}
\label{eq:SI31}
\begin{split}
  H =&\,  \frac{\omega_{\mathrm{q}}(t)}{2} \, \sigma_\mathrm{z} + \omega_\mathrm{m}(t) b^\dagger b \\
     & +  g_\mathrm{1}(t) (b^\dagger + b) \sigma_\mathrm{x} 
       +  g_\mathrm{2}(t) (b^\dagger + b)^2 \sigma_\mathrm{z}
\end{split}
\end{equation}
will be time-dependent. Up to second order in $\phi_\mathrm{b0}$, and
first order in $x_\mathrm{ZPF}/\xi$, it
is possible to write them as

  \begin{alignat}{3}
    \label{eq:SI32}
   &\omega_{\rm q}(t)  &&\simeq 2\left(\omega_{\rm p0}-E_{\rm C}-\frac{\phi_{b0}^2 \omega_{\rm p0}}{16} \frac{1+\cos\left(2 \bar{\omega} t\right)}{2}\right) \nonumber \\ &         && = \bar{\omega}_{\rm q} - \delta_{\rm q} \cos\left(2 \bar{\omega} t\right) \nonumber \\
    &g_{\rm 1}(t) && \simeq \bar{g}_1 \cos \left(\bar{\omega} t\right) \\
    &\omega_{\rm m}(t) && \simeq \omega_{\rm m}(0)  \nonumber\\
    &g_{2}(t) && \simeq 0  \nonumber
  \end{alignat}
where
$\bar{\omega}_{\rm q}=\left. \omega_{\rm q}\right|_{\phi_{\rm b}=0}-\delta_{\rm q}$,
$\delta_{\rm q}=\frac{\phi_{b0}^2 \omega_{\rm p0}}{16}$,
$\bar{g}_1= E_\mathrm{J} (\frac{\ecz}{E_\mathrm{J}})^{1/4}\left(\sqrt{\frac{\ecz}{E_\mathrm{J}}}-1 \right)\left(\frac{x_\mathrm{ZPF}}{\xi}\right)\phi_{\rm b0} $.
With these approximations, we move to a rotating frame both for the
qubit and the mechanical mode with the unitary transformation 
\begin{align}
  \label{eq:SI33}
  U=\exp\left[-\frac{i}{2} \left\{\bar{\omega}_{\rm q}t -
                                    \frac{\delta_\mathrm{q}}{2\bar{\omega}} \sin \left( 2 \bar{\omega t}\right)
                           \right\} \sigma_{\rm z} -
                           i \omega_\mathrm{m} b^\dagger b
         \right].
\end{align}
In this frame the transformed Hamiltonian
$H' =  U^\dagger H U - U^\dagger \dot{U}$ can be written as
\begin{align}
  \label{eq:SI34}
  H' &= \bar{g}_1 \cos \left(\bar{\omega} t\right) \left(b^\dagger e^{i \omega_\mathrm{m} t} + b e^{-i \omega_\mathrm{m} t}\right) \left[\cos \alpha(t) \sigma_\mathrm{x} + \sin \alpha(t) \sigma_\mathrm{y} \right] \nonumber \\
     &= \bar{g}_1 \cos \left(\bar{\omega} t\right) \left(b^\dagger e^{i \omega_\mathrm{m} t} + b e^{-i \omega_\mathrm{m} t}\right) \left(e^{-i \alpha(t)}\sigma_+ + e^{i \alpha(t)}\sigma_- \right) 
\end{align}
with $\sigma_x=\sigma_++ \sigma_-$,
$\sigma_y=i\left(\sigma_+- \sigma_-\right)$, and $\alpha(t)= \bar{\omega}_{\rm q}t - \frac{\delta_\mathrm{q}}{2\bar{\omega}} \sin \left( 2 \bar{\omega t}\right)$.
Considering the relation
\begin{align*}
  \exp \left[i z \sin \theta \right]=\sum_{k=-\infty}^{\infty} e^{i k \theta} J_k(z) ,
\end{align*}
where $J_k(z)$ is the n-th order Bessel function, we can write $H'$ as
\begin{align}
  \label{eq:SI35}
  H' = \bar{g}_1 \cos \left(\bar{\omega} t\right)
  \left(b^\dagger e^{i \omega_\mathrm{m} t} + b e^{-i \omega_\mathrm{m} t}\right)
  \left(
     e^{i \bar{\omega}_\mathrm{q} t} \left[ J_0\left(\frac{\delta_\mathrm{q}}{2 \bar{\omega}}\right) + J_1\left(\frac{\delta_\mathrm{q}}{2 \bar{\omega}}\right)e^{2 i \bar{\omega} t}\dots \right]  \sigma_+ + \mathrm{h.\,c.}\right) .
\end{align}

For $\bar{\omega}=\bar{\omega}_\mathrm{q}-\omega_\mathrm{m}$, invoking
the rotating wave approximation and neglecting non-resonant terms,
Eq.~\eqref{eq:SI35} can be written as
\begin{align}
  \label{eq:SI36}
   H' = \bar{g}_1 J_0\left(\frac{\delta_\mathrm{q}}{2 \bar{\omega}}\right) \left( b^\dagger \sigma_- + \mathrm{h.\,c.}\right).
\end{align}
With
$\bar{g}_1 J_0\left(\frac{\delta_\mathrm{q}}{2 \bar{\omega}}\right)=g_{\rm sw}$
this is the expression given in Eq.~(7) of the main text.

% *** continue with the expansion and the discussion.

% \begin{align*}
%   \exp\left[i \left(\bar{\omega}_\mathrm{q} t - \frac{\delta_\mathrm{q}}{2 \bar{\omega}} \sin 2 \bar{\omega} t \right)\right]=  \exp\left[i \left(\bar{\omega}_\mathrm{q} t] \sum_{k=-\infty}^{\infty} e^ 
% \end{align*}

% The \nocite command causes all entries in a bibliography to be printed out
% whether or not they are actually referenced in the text. This is appropriate
% for the sample file to show the different styles of references, but authors
% most likely will not want to use it.
%\nocite{*}
\bibliographystyle{apsrev_abb}
\bibliography{references_X2mon_supplementary.bib}% Produces the bibliography via BibTeX.